\documentclass[journal]{IEEEtran}
\usepackage{subcaption}
\usepackage{graphicx}
\usepackage{amsmath,amssymb,bm,mathrsfs,xfrac,mathtools,mathrsfs,amsthm, nccmath}
\usepackage{color, colortbl}
\usepackage[export]{adjustbox}
\usepackage{tikz}
\usetikzlibrary{positioning}
\usetikzlibrary{shapes.geometric, arrows}
\usepackage{longtable}
\usepackage{lscape}
\usetikzlibrary{calc,shapes,arrows,automata,positioning} %flowchart
\usepackage{float}
\usepackage{cite}
\usepackage[hyphens]{url}
\usepackage[hidelinks]{hyperref}
\usepackage{soul}
\usepackage{blindtext}
\usepackage{multirow}
\usepackage{comment}
\usepackage{forest}
\usepackage{pgf}
\usepackage{pifont}
\usepackage{dashrule}
\usepackage{booktabs}
\definecolor{NREL1}{RGB}{000,166,222}

\ifCLASSINFOpdf
\else
\fi

\begin{document}
\bstctlcite{IEEEexample:BSTcontrol}

\title{A Review on Symbolic Regression in Power Systems: Methods, Applications, and Future Directions%Symbolic Regression: Learning Parsimonious Power System Models from Data
}
\author{Amir Bahador Javadi, \IEEEmembership{Student Member, IEEE,} and Philip~Pong,~\IEEEmembership{Senior~Member,~IEEE}
%and Amin~Kargarian,~\IEEEmembership{Senior~Member,~IEEE}
\thanks{This material is based upon work supported by the National Science Foundation under Grant Number 2328241. Any opinions, findings, and conclusions or recommendations expressed in this material are those of the authors and do not necessarily reflect the views of the National Science Foundation.}
\thanks{ 
A.B. Javadi, and P. Pong are
with the Helen and John C. Hartmann Department of Electrical and Computer Engineering, New Jersey Institute of Technology, Newark, NJ 07102 USA (e-mail: \{aj772, philip.pong\}@njit.edu).}}

%A. Kargarian are with the Department of Electrical and ComputerEngineering, Louisiana State University, Baton Rouge, LA 70803 USA (e-mail: ajavad2@lsu.edu; kargarian@lsu.edu).

% \markboth{IEEE TRANSACTIONS ON POWER SYSTEMS,~Vol.~, No.~, ~2021}
% {Shell \MakeLowercase{\textit{et al.}}: Bare Demo of IEEEtran.cls for Journals}

\maketitle

\begin{abstract}

As power systems evolve with the increasing integration of renewable energy sources and smart grid technologies, there is a growing demand for flexible and scalable modeling approaches capable of capturing the complex dynamics of modern grids. This review focuses on symbolic regression, a powerful methodology for deriving parsimonious and interpretable mathematical models directly from data. The paper presents a comprehensive overview of symbolic regression methods, including sparse identification of nonlinear dynamics, automatic regression for governing equations, and deep symbolic regression, highlighting their applications in power systems. Through comparative case studies of the single machine infinite bus system, grid-following, and grid-forming inverter, we analyze the strengths, limitations, and suitability of each symbolic regression method in modeling nonlinear power system dynamics. Additionally, we identify critical research gaps and discuss future directions for leveraging symbolic regression in the optimization, control, and operation of modern power grids. This review aims to provide a valuable resource for researchers and engineers seeking innovative, data-driven solutions for modeling in the context of evolving power system infrastructure.

\end{abstract}

\begin{IEEEkeywords}
ARGOS, automatic regression for governing equations, deep symbolic regression, model identification, SINDy, sparse identification of nonlinear dynamics, sparse regression, symbolic regression, system identification.
\end{IEEEkeywords}

\IEEEpeerreviewmaketitle

\section{Introduction}

\IEEEPARstart{S}{ymbolic} regression (SR) is a \emph{process of discovering parsimonious mathematical expressions from datasets}.

The term SR was first introduced by John R. Koza in \cite{10.5555/892491} to address the challenge of not only determining the numerical coefficients of a predefined function, as in traditional regression methods, but also discovering the underlying functional form that best fits the data. Unlike linear or polynomial regression, where the model's structure is selected beforehand, SR searches for both the optimal mathematical expression and its corresponding parameters. This approach is particularly valuable when the most appropriate type of function for a given dataset is unknown, allowing for the exploration of both polynomial and nonpolynomial families of functions. 

This area was extended by the introduction of sparse identification of nonlinear dynamics (SINDy) by Steven L. Brunton et al. \cite{Brunton2016}, which applied the method to dynamical systems for identifying governing equations from various datasets, including examples from fluid dynamics, chemical kinetics, and neuroscience. 
In 2019, the concept of neuro-symbolic artificial intelligence emerged, emphasizing the fusion of deep learning and symbolic artificial intelligence techniques. Neuro-symbolic approaches aim to combine the statistical power of deep learning with the reasoning capabilities of SR, enabling machines to learn and reason in a more human-like manner \cite{serafini2019neuro}. 
Researchers have developed novel architectures and algorithms specifically tailored for deep SR tasks \cite{karazeev2019deep}. These models typically involve neural network components for feature extraction and representation learning, combined with SR techniques for generating interpretable formulas or expressions. 
Recently, Egan et al. \cite{egan2024automatically} presented automatic regression for governing equations (ARGOS), a new approach integrating signal denoising, sparse regression, and bootstrap confidence intervals to automatically uncover system dynamics.

SR, linear regression, and system identification are distinct approaches to modeling data and systems. SR is a powerful tool at the intersection of applied mathematics and machine learning, providing deep insights into the underlying relationships within complex, nonlinear systems where the governing equations are unknown, without requiring a predefined model structure \cite{vladislavleva2009symbolic}. Linear regression models the relationship between a dependent variable and one or more independent variables, assuming a linear form. This makes it simple and interpretable, but limited to linear relationships \cite{montgomery2012introduction}. System identification focuses on building dynamic models from input-output data by fitting a predefined structure, such as state-space models or transfer functions. It is commonly used in control engineering for dynamic systems \cite{ljung1998system}. A comparison of SR, linear regression, and system identification is provided in Table \ref{tab:comparison}, highlighting key attributes such as flexibility and interpretability.

\begin{table}[htbp]
\caption{Comparison of SR, linear regression, and system identification.}
\centering
\resizebox{0.5\textwidth}{!}{
\begin{tabular}{ccc}
\toprule
\textbf{Method}                   & \textbf{Flexibility}           & \textbf{Interpretability} \\ \midrule
\textbf{Symbolic regression}  & Highly flexible                & Highly interpretable      \\ 
\textbf{Linear regression}         & Limited to linear models     & Highly interpretable      \\ 
\textbf{System identification}     & Restricted by predefined models & Moderately interpretable  \\ \bottomrule
\end{tabular}
}
\label{tab:comparison}
\end{table}

%The main differences between these approaches lie in their assumptions and flexibility. SR is highly flexible and interpretable, especially when it identifies simple expressions. In contrast, linear regression is restricted to linear relationships, and system identification relies on predefined model structures, making it less flexible but well-suited for dynamic systems \cite{bongard2007automated}. The choice of method depends on the system's complexity and the desired level of interpretability.
%Various methods have emerged within the realm of SR to tackle the challenge of discovering equations from datasets. 

As power systems undergo rapid transformation with the increasing integration of renewable energy sources and smart grid technologies, these changes demand more flexible and scalable modeling approaches. Traditional methods may not always capture the nuances of these new grid dynamics, making SR an ideal tool to develop accurate, interpretable models without requiring deep physical insights or assumptions \cite{5546958}. A recent publication by Makke and Chawla \cite{makke2024interpretable} comprehensively reviews various SR methods, yet there remains a notable lack of dedicated literature focusing on their specific applications within power systems. A focused review in this area would fill this gap and provide a valuable resource for researchers and engineers seeking tailored solutions for modeling, optimization, and control in modern power grids.

%In this review paper, we aim to examine SR methods and their applications within the context of power systems. We will present a comparative analysis of the most practical SR methods, particularly in relation to the single machine infinite bus system. Furthermore, this study will provide a critical assessment of the strengths and limitations inherent in each method, thereby identifying potential research gaps and applications. 
This paper is a focused review of SR methods, with particular emphasis on their applications in the field of power systems. We examine key SR methods namely SINDy, ARGOS, and deep SR, analyzing their advantages, limitations, and suitability for modeling power system dynamics, especially through case studies of the single machine infinite bus system, grid-following, and grid-forming control modes of an inverter. Furthermore, this paper aims to identify key challenges and research gaps, while laying the groundwork for further investigation and exploration in this field. Additionally, we have carefully selected a diverse range of articles published in reputable journals and conferences, with a specific focus on renewable energy and converter-based resources, to ensure a comprehensive and high-quality review. Since power systems are predominantly modeled using ordinary differential equations and differential-algebraic equations, rather than purely algebraic or partial differential equations, our focus will also encompass applications relevant to these mathematical frameworks.

The paper is not an exhaustive review of SR across all scientific disciplines. While many domains, including physics and biology, have utilized SR for various modeling tasks, this review specifically narrows its scope to power systems applications. Additionally, this review does not aim to provide an in-depth exploration of machine learning approaches beyond SR. It also does not offer a comprehensive comparison with black-box machine learning techniques like deep neural networks, which are often less interpretable but powerful in high-dimensional datasets.

%This review seeks to bridge the gap in the literature by focusing on SR's use in power system applications and providing practical insights for researchers and engineers working in this domain.

%To comprehensively explore the enhancement of performance and innovation in power systems through SR, this review aims to address existing gaps by thoroughly examining recent advancements in various approaches within this crucial domain. As power systems become increasingly complex, the demand for efficient data management and storage also rises. This underscores the necessity of comparing all available SR approaches, including genetic programming, SINDy, and deep SR, which can effectively handle these challenges.
%Table 1 illustrates the increasing complexity of power system model and the growing volume of data per day, alongside the evolution of SR methods from the 1990s to the present, with a prediction of this growing trend for the next decade. 
%This paper seeks to elucidate the development, evolution, and practical implementation of SR methods within the domain of power systems, aiming to identify key challenges and research gaps while laying the groundwork for further investigation and exploration in this field. By highlighting these gaps, the paper not only sets the stage for future studies but also suggests directions that could enhance the application of SR in power systems. 
%In addition, this review highlights the potential of SR in the sparse modeling, particularly in the context of renewable energy and converter-based resources.

The remaining sections of this paper are structured as follows: Section \ref{sec:2} delves into SR methods within the domain of power systems, providing a comprehensive review of practical approaches adopted across various domains. Here, methodologies such as SINDy, ARGOS, and deep SR are discussed briefly. Section \ref{sec:3} explores the applications of these methods within this area, examining the widespread applications of SINDy. Additionally, the section delves into the relatively recent technique of deep SR. Section \ref{sec:4} presents a comparative analysis on three simple case studies in power systems for a better understanding and insights towards all approaches. In section \ref{sec:5}, the authors offer their perspectives on the topic, providing insights into potential future directions, and critically analyze the strengths and limitations of each method, discussing how these challenges could pave the way for future research endeavors. Finally, Section \ref{sec:6} encapsulates the entire work, providing a conclusive summary of the findings and discussions presented throughout the paper.

\section{Overview of SR Methods}

\label{sec:2}
\begin{figure*}[ht]
    \centering
    \includegraphics[width=1\textwidth]{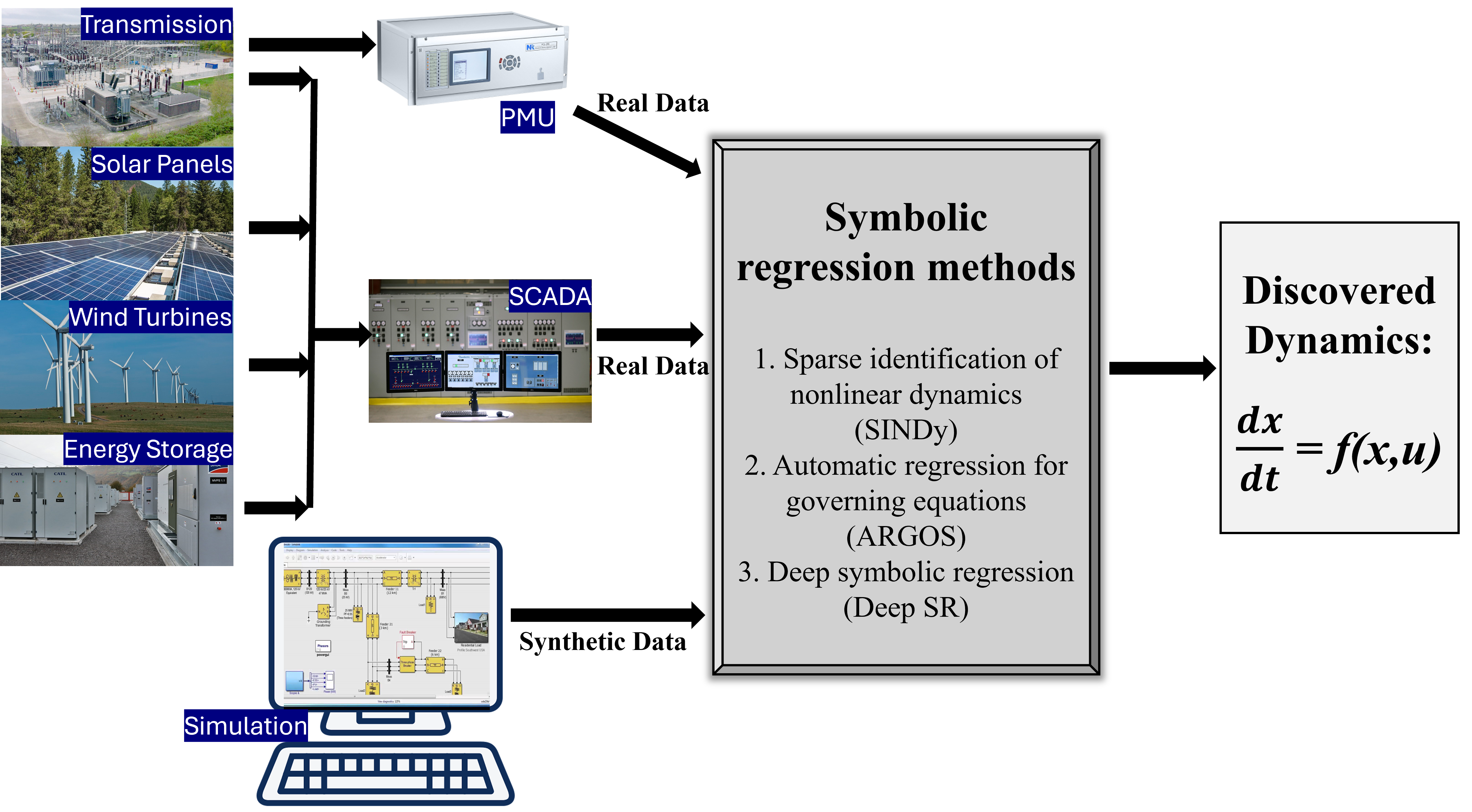}
    \caption{Diagram representing the workflow for model identification using symbolic regression methods, from collecting real or synthetic data to identifying the system in terms of mathematical expressions.}
    \label{fig:high_level}
\end{figure*}

This section provides a concise overview of the foundational aspects pertinent to each method. This will enhance comprehension and facilitate the utilization of these concepts within power systems applications. Initially, a succinct examination of SINDy will be reviewed, outlining the procedural steps involved in arriving at the ultimate solution. Following that, ARGOS will be explored, elucidating its principles and methods. Lastly, deep SR will be introduced, providing insight into its functionality and relevance. The systematic workflow of symbolic regression methods, starting from data collection (real or synthetic) to deriving mathematical expressions that define the identified system is illustrated in Fig. \ref{fig:high_level}. 

Consider a dynamical system:
\begin{align*}
\begin{cases}
\dot{\mathbf{x}}(t) &= f(\mathbf{x}(t),\mathbf{u}(t)), \\
\mathbf{x}(0) &= \mathbf{x}_0
\end{cases}
\end{align*}

where \( \mathbf{x}(t) \in \mathbb{R}^n \) is the state vector,
\( \mathbf{u}(t) \in \mathbb{R}^m \) is the control input vector, $\mathbf{x}_0$ is the initial condition of the system, \( f(\mathbf{x}, \mathbf{u}) \) is the unknown function governing the system dynamics.

\subsection{Sparse Identification of Nonlinear Dynamics (SINDy)}

In SINDy, the goal is to identify \( f(\mathbf{x}, \mathbf{u}) \) in a symbolic form while ensuring the solution is sparse \cite{doi:10.1073/pnas.1906995116}. The concept of sparsity, which refers to the idea that many natural phenomena can be described by a small number of significant terms in a mathematical model, gained prominence in various fields including signal processing and compressed sensing \cite{4472240}. This method was developed as a way to leverage sparsity in the context of identifying nonlinear dynamical systems. It aimed to find the simplest model that could describe the observed dynamics, exploiting the fact that in many systems, only a few terms are dominant while others can be neglected \cite{schaeffer2017learning}. 

In order to address that, a large set of candidate functions \( \Theta(\mathbf{x}, \mathbf{u}) \) should be constructed, which includes potential nonlinear terms of \( \mathbf{x}(t) \) and \( \mathbf{u}(t) \) \cite{doi:10.1126/sciadv.1602614}. This library of functions may consist of virtually any function, e.g. monomials, polynomials, trigonometric, rational, etc.
%\[
%\Theta(\mathbf{x}, \mathbf{u}) = [1, x_1, x_2, u_1, u_2, x_1^2, x_1 x_2, u_1 x_1, \dots]^\top.
%\]
The system dynamics can then be approximated as a linear combination of these candidate functions:
\[
\dot{\mathbf{x}}(t) \approx \Theta(\mathbf{x}, \mathbf{u}) \xi,
\]
where \( \xi \) is the vector of coefficients that determines which terms from the library are used to describe the dynamics.

To obtain a sparse model, the following optimization problem should be solved \cite{doi:10.1137/18M1189828}:
\[
\min_{\xi} \|\dot{\mathbf{x}}(t) - \Theta(\mathbf{x}, \mathbf{u}) \xi\|_2 + \lambda \|\xi\|_0,
\]
where \( \|\dot{\mathbf{x}}(t) - \Theta(\mathbf{x}, \mathbf{u}) \xi\|_2 \) represents the least-squares error, \( \|\xi\|_0 \) is the \( \ell_0 \)-norm enforcing sparsity by minimizing the number of non-zero elements in \( \xi \), although in practice, the \( \ell_0 \)-norm is often replaced by the \( \ell_1 \)-norm for computational tractability \cite{51791361-8fe2-38d5-959f-ae8d048b490d}. Additionally, \( \lambda \) controls the trade-off between accuracy and sparsity.

When incorporating control inputs, \( \mathbf{u}(t) \), they are treated similarly to state variables but represent external forces acting on the system. The library \( \Theta(\mathbf{x}, \mathbf{u}) \) includes interaction terms between state and control variables, allowing the model to capture the control input’s effect on system dynamics \cite{brunton2022data}. %Sparse SR encompasses two primary methodologies: sparse identification of nonlinear dynamics (SINDy) \cite{Brunton2016} and the automatic regression for governing equations (ARGOS) \cite{egan2024automatically}. 

SINDy leverages techniques from compressed sensing and sparse regression to identify the governing equations of dynamical systems directly from data \cite{ljung1999system,1614066}. By exploiting sparsity-promoting techniques, SINDy can uncover interpretable mathematical models from high-dimensional data. %Prior to SINDy, various methods for identifying dynamical systems from data, such as system identification techniques and model order reduction methods, were computationally intensive or limited in applicability \cite{sarkka2013bayesian,tipping2001sparse}.
%This method quickly gained recognition and was widely adopted across various scientific disciplines. Its simplicity, efficiency, and ability to handle incomplete data made it particularly appealing to researchers working with complex systems where traditional modeling approaches were challenging. 
Since its introduction, this technique has undergone further development and refinement. Researchers have explored extensions and variations of the original algorithm to improve its performance in different contexts and to address specific challenges \cite{fasel2022ensemble,hirsh2022sparsifying,kaheman2022automatic,mangan2017model,brunton2016sparsecontrol}. SINDy has also been integrated with machine learning techniques, such as deep learning and reinforcement learning, to enhance its capabilities and tackle more complex problems \cite{baumeister2018deep}.
 
\subsection{Automatic regression for governing equations (ARGOS)}
The ARGOS \cite{egan2024automatically} method offers a robust and efficient approach to identifying the underlying dynamics of a system from noisy and limited data. Building on sparse regression techniques, ARGOS integrates signal smoothing, automatic feature selection, and statistical inference to automatically discover nonlinear differential equations that govern system behavior. By leveraging bootstrapped confidence intervals and sparse regression methods, ARGOS aims to overcome the traditional challenges in model discovery, particularly in the presence of high noise and complex dynamics, while ensuring the interpretability of the resulting models. In ARGOS, the dynamics of the system is approximated by using a symbolic representation:
\begin{equation*}
    \dot{\mathbf{x}}_j \approx \theta^{T}_F (\mathbf{x}) \beta_j, \quad j = 1, \dots, n
\end{equation*}
where $\theta_F (\mathbf{x})$ is a feature matrix constructed from \( \mathbf{x} \) and nonlinear combinations of its terms, \( \beta_j \) is a sparse vector of coefficients to be identified. The feature matrix \( \Theta(\mathbf{x}) \) using monomials up to a certain degree \( d \), which can also include nonlinear functions, should be constructed. After that, the system is expressed in matrix form as:
\begin{equation*}
    \dot{\mathbf{x}} = \Theta(\mathbf{x}) B + E
\end{equation*}
where \( B \) is a matrix of coefficients \( \beta_j \), and \( E \) is the residual matrix representing model errors. To identify sparse models, the least absolute shrinkage and selection operator is used \cite{51791361-8fe2-38d5-959f-ae8d048b490d}:
\begin{equation*}
    \hat{\beta} = \arg\min_\beta \left\{ \|\dot{\mathbf{x}} - \Theta(\mathbf{x})B\|_2^2 + \lambda \sum_k w_k |\beta_k| \right\}
\end{equation*}
where \( w_k \) are the weights for the adaptive least absolute shrinkage and selection operator \cite{zou2006adaptive}, and \( \lambda \) is the regularization parameter. After the initial sparse regression step, the design matrix is trimmed to retain only selected terms. Regression is repeated to refine the coefficients and obtain unbiased estimates. Finally, bootstrap sampling is applied to generate multiple estimates of the coefficients, allowing the construction of confidence intervals:
\begin{equation*}
    \text{Confidence interval for } \beta_k: \left[ \hat{\beta}_{k,\text{low}}, \hat{\beta}_{k,\text{up}} \right]
\end{equation*}
Variables whose confidence intervals do not include zero are considered significant and will be used for discovering the dynamics of the system. %Through its automated process, ARGOS significantly improves the accuracy and efficiency of discovering dynamical systems, outperforming traditional methods like SINDy, particularly in higher-dimensional and noisy environments. The ability to automatically tune parameters, reduce overfitting, and provide confidence intervals makes ARGOS a powerful tool for researchers dealing with complex systems where data is abundant but mathematical models are difficult to construct manually. With its comprehensive framework, ARGOS paves the way for more widespread and practical application of system identification techniques across various scientific disciplines

%ARGOS-RAL \cite{li2024automating}, an extension of the ARGOS framework, uses sparse regression with recurrent adaptive lasso to automatically identify partial differential equations from minimal prior knowledge. The method automates derivative calculations, constructs a candidate library, and estimates sparse models. Tested under varying noise levels and sample sizes, ARGOS-RAL proves robust in handling noisy and unevenly distributed data, outperforming sequential threshold ridge regression in most cases. This approach highlights the potential of combining statistical methods, machine learning, and dynamical systems theory to streamline scientific modeling. ARGOS and ARGOS-RAL are novel methodologies that have yet to be implemented in power systems. However, we include them in this discussion due to their potential for accurately capturing power system dynamics in noisy environments. These two methods could be applied to power system applications, allowing for an assessment of their capabilities in elucidating the underlying dynamics of such systems.

ARGOS is a novel methodology that has not yet been implemented in power systems. However, we include it in this discussion due to its potential to accurately capture power system dynamics in noisy environments. This method could be applied to various power system applications, enabling an assessment of its effectiveness in uncovering the underlying system dynamics.

The most significant difference between ARGOS and SINDy lies in the degree of automation and robustness in handling noisy data. ARGOS extends SINDy by incorporating automated parameter tuning, particularly in denoising the data using the Savitzky-Golay filter \cite{savitzky1964smoothing}, where ARGOS optimally selects the filter parameters, while SINDy requires manual tuning. ARGOS also employs bootstrap sampling to generate confidence intervals for the identified model parameters, providing statistical guarantees that SINDy lacks. Furthermore, ARGOS introduces a more efficient and adaptive model selection process through the adaptive lasso and the Bayesian information criterion, improving its accuracy and computational performance in higher-dimensional and noisy systems. In contrast, SINDy is more prone to errors in such cases due to manual thresholding and difficulty in handling multicollinearity in the feature matrix. Nevertheless, SINDy can discover the underlying dynamics of the system much faster than ARGOS, making it a potential tool for real-time and online system modeling. These improvements make ARGOS more suitable for complex real-world systems with medium noise levels and larger datasets.

\subsection{Deep SR}

Deep SR is an innovative fusion of deep learning methodologies such as neural networks with traditional SR methods.
%Deep SR represents a modern approach that combines the power of deep learning with SR. 
By integrating neural networks with SR methods, deep SR endeavors to learn both the numerical representations and symbolic structures of underlying equations, facilitating the discovery of complex relationships in data while maintaining human-interpretable forms \cite{petersen2019deep}. This method has found applications in various domains, including physics, biology, chemistry, finance, and engineering, and has been used for tasks such as equation discovery, function approximation, system identification, and modeling complex phenomena \cite{ellis2018neuro}. %Researchers have also begun exploring ways to combine deep learning with SR techniques to leverage the strengths of both approaches \cite{RAISSI2019686}. This integration aimed to harness the representational power of neural networks for capturing complex patterns in data while also generating human-interpretable symbolic expressions.
%This technique aims to uncover symbolic expressions directly from data, making it particularly relevant for SR in learning parsimonious models from data. 

By blending the expressive capabilities of deep neural networks to discern intricate patterns within the data, it emphasizes generating interpretable symbolic expressions, which not only capture complex data patterns but also offer insights into the fundamental mechanisms governing the model of the system under study. Here is an outline of the steps and procedures typically involved in this method:

\begin{enumerate} 

 \item \textit{Data Preprocessing}: This step involves preparing the data for analysis, including cleaning, normalization, and feature selection if necessary. It is crucial to ensure that the data is in a suitable format for the regression task \cite{doi:10.1126/science.1165893}. 
 \item \textit{Deep Learning Model}: A deep neural network architecture is chosen to learn the underlying patterns in the data. This could be a convolutional neural network, recurrent neural network, or a combination of various layers depending on the nature of the data and the problem at hand \cite{petersen2019deep,NEURIPS2020_c9f2f917}.
 \item \textit{SR}: SR techniques are employed to discover mathematical expressions that symbolically represent the learned relationships within the data. SR algorithms typically involve searching through a space of mathematical expressions using evolutionary algorithms, genetic programming, or other optimization techniques \cite{mundhenk2021symbolic,10.5555/3600270.3602733}.
 \item \textit{Fitness Evaluation}: The quality of symbolic expressions is evaluated using a fitness function that measures how well the expressions fit the data. This evaluation is typically based on criteria such as accuracy, simplicity, and generalization performance \cite{pmlr-v139-biggio21a}.
 \item \textit{Optimization}: The SR algorithm iteratively refines the set of candidate expressions by selecting and modifying them based on their fitness scores. This process continues until satisfactory expressions are found or a predefined stopping criterion is met \cite{virgolin2022symbolic}.

\end{enumerate}

%Traditional SR methods primarily rely on genetic programming to evolve symbolic expressions through operations like mutation and crossover. While these methods offer high interpretability and flexibility, they often suffer from slow convergence, limited scalability, and inefficient exploration of the solution space, particularly for high-dimensional or noisy datasets (Koza, 1992; Vladislavleva et al., 2009). In contrast, deep symbolic regression (Deep SR) leverages advances in deep learning, such as recurrent neural networks or transformer architectures, in conjunction with reinforcement learning or Bayesian optimization to guide the search for equations more efficiently. This modern approach enhances the scalability and data efficiency of symbolic regression while maintaining interpretability (Sahoo et al., 2018; Cranmer et al., 2020).

%Fig. \ref{fig:dsr_flowchart} represents the sequence of all steps in this method, highlighting that 
%Deep learning model step and SR stage can be executed concurrently to reduce the time required for capturing the dynamics from data. 
Given our objective to scrutinize the applications of these methodologies within power systems domains, it is imperative to direct attention to the following references for a thorough investigation and enhanced comprehension \cite{doi:10.1126/sciadv.aay2631,NEURIPS2020_33a854e2,kim2020integration,zhang2023deep,keren2023computational,d2023odeformer,du2024large,biggio2020seq2seq,kamienny2022end,shojaee2023transformer,shojaee2024llm}.

\section{Applications of SR Methods}
\label{sec:3}

This section provides a thorough examination of various applications within power systems. It begins by surveying the use of a well-established methodology called SINDy in this domain, then explores the applications of deep SR.

\subsection{Sparse Identification of Nonlinear Dynamics (SINDy)}

Currently, SINDy has gained significant popularity among researchers for discerning the mathematical patterns within datasets. This approach has particularly garnered attention for its efficacy in deciphering complex datasets and elucidating the behavior of systems. For instance, within the area of power systems, this method has found widespread application in refining and pinpointing models based on data. In \cite{9237125,stankovic2020data}, researchers propose a methodology tailored for identifying and modeling power systems utilizing data-driven techniques, with a specific focus on synchronous generator models. This method integrates SINDy with other methodologies to construct models that adeptly capture system behavior, thereby rendering them suitable for real-time applications, particularly in response to dynamic system changes. 
%Additionally, Javad Khazaei and Faegheh Moazeni \cite{khazaei2023model} have pioneered a methodology aimed at modeling distributed energy resources through data-driven techniques rather than relying on intricate physics-based models. Leveraging SINDy in conjunction with Koopman theory, their framework adeptly captures the dynamics of these resources, streamlining control design processes. Through simulations, this approach demonstrates promising outcomes, indicating its potential to offer efficient and accurate representations of distributed energy resources behavior while facilitating simplified control strategies.

In the quest to comprehend and model the intricate dynamics inherent in microgrid systems, researchers employ this method to unravel the behavior of microgrids amid disturbances, particularly as their inertia diminishes due to the integration of distributed energy resources. This algorithm proves invaluable in identifying the nonlinear dynamics characterizing the system, furnishing a dependable tool for analyzing transient dynamics in the absence of an accurate system model \cite{nandakumar2022sparse}. Taking a stride forward, the same authors venture into the domain of modularized SINDy, known as M-SINDy, elucidated in \cite{nandakumar2023data}. This innovative approach entails deconstructing the system into more manageable components and utilizing pseudo-states to encapsulate their interactions. Through this process, modularized SINDy showcases its efficacy in discerning the governing equations governing various subsystems within the microgrid. Such a methodological approach not only augments stability but also fortifies the reliability of microgrids, thereby paving promising pathways for enhancing their operational efficiency and resilience.

The area of model identification from datasets has expanded its scope to encompass power converters, as highlighted in \cite{hosseinipour2023sparse,hosseinipour2023data}. These papers introduce a novel approach leveraging sparse regression techniques to unravel the behavior of power converters within microgrids. By scrutinizing data obtained during converter operation, this method adeptly constructs precise models crucial for control strategies and stability assessments, all without necessitating continual system perturbations. Through rigorous simulations and frequency analyses, the effectiveness of this approach is vividly demonstrated, underscoring its potential to significantly enhance the understanding and management of power converters in microgrid systems. This approach proves valuable not only in deciphering power converter behavior but also in understanding load model dynamics. In \cite{10194488}, a fresh methodology emerges, merging SR with sparse dictionary learning to refine models of power system loads. The overarching goal of this method is to elevate the precision and utility of load models within practical settings by amalgamating diverse mathematical functions and subjecting them to various fault scenarios for comprehensive testing. By doing so, it endeavors to enhance the accuracy and efficacy of load modeling, thereby offering promising strides towards more robust and adaptable real-world applications. 

This approach finds application in yet another crucial domain: the control of power systems. Demonstrating its versatility, \cite{tran2023sparse,shadaei2024data} showcase its deployment in distinct contexts, namely the recovery of model predictive control for long-term voltage stability and power allocation among inverter-based resources, respectively. In \cite{tran2023sparse}, the authors delve into the utilization of model predictive control systems to orchestrate distributed energy resources and grid controllers, particularly focusing on voltage restoration post-emergencies. This approach, integrating SINDy to forecast voltage evolution, coupled with an adaptive model predictive control within a centralized controller, notably reduces voltage recovery time while minimizing control interventions. In alignment with this, \cite{shadaei2024data} introduces a novel approach named non-linear model predictive control tailored for managing energy resources within smart grids. By relying on measurements rather than models, it facilitates precise power distribution among resources, thereby enhancing voltage and frequency control. Extensively trialed on a microgrid, this method exhibits its resilience and suitability compared to traditional model predictive control strategies, underscoring its potential to revolutionize energy resource management in modern grid systems. 

When discussing the management of distributed energy resources, SINDy emerges as a pivotal tool for modeling and controlling these resources within smart grid contexts, obviating the necessity for intricate system dynamics or extensive historical data. Its application extends across a spectrum of grid scenarios, ranging from black-start situations to the integration of distributed energy resources into weak grids and microgrids, as elucidated in \cite{khazaei2023data,khazaei2023data1}. This approach proves highly effective in safeguarding stability and optimizing performance across diverse grid environments, offering a streamlined yet robust solution for distributed energy resources management within the smart grid framework. As the integration of inverter-based resources escalates within the power grid, the system's complexity surges, necessitating a pragmatic toolkit to govern its mathematical behavior. Addressing this imperative, authors in \cite{en17030711} deliberate on the intricacies posed by modern power grids replete with renewable energy sources, underscoring the indispensability of advanced algorithms for precise system modeling and optimization. Employing SINDy, they endeavor to pinpoint crucial features and delineate system-level nonlinearity via an innovative index. Their exploration reveals a discernible shift in dynamics contingent upon alterations in power sources, signifying the critical role of such methodologies in navigating the evolving landscape of power grid complexities.

Power system parameter estimation emerges as a focal point for the application of SINDy. In \cite{su14042051,9862464}, authors delve into the utilization of this approach for parameter estimation within power grids, employing a fusion of physics-based principles and machine learning techniques. Their investigation unveils the faster performance of the SINDy algorithm in comparison to an alternative method known as physics-informed neural networks, particularly excelling in accurately estimating power grid parameters, notably in systems characterized by slow dynamics. Moreover, the computational efficiency of SINDy renders it well-suited for real-time applications, accentuating its potential as a robust tool in the realm of power system parameter estimation.

\begin{table*}[h!]
\begin{center}
\footnotesize
\captionsetup{skip=-0.5pt}
\caption{\raggedright Applications of SINDy in power systems, highlighting their key features and references.}
\label{tab:sindy}
\begin{tabular}{>{\raggedright\arraybackslash}p{0.2\textwidth}>{\raggedright\arraybackslash}p{0.25\textwidth}>{\raggedright\arraybackslash}p{0.35\textwidth}>{\centering\arraybackslash}p{0.09\textwidth}}
\hline \hline
\textbf{Methodology} & \textbf{Application} & \textbf{Key Features} & \textbf{References} \\ \hline
SINDy with data-driven techniques & Modeling synchronous generators in power systems & Constructs models capturing system behavior suitable for real-time applications, especially dynamic changes. & \cite{9237125,stankovic2020data} \\ \hline
%SINDy with Koopman theory & Modeling distributed energy resources & Efficiently captures distributed energy resources dynamics, streamlining control design processes. & \cite{khazaei2023model} \\ \hline
SINDy & Analyzing microgrid dynamics & Identifies nonlinear dynamics and provides a reliable tool for analyzing transient dynamics. & \cite{nandakumar2022sparse} \\ \hline
Modularized SINDy (M-SINDy) & Enhancing microgrid stability and reliability & Deconstructs systems into components using pseudo-states to encapsulate interactions. & \cite{nandakumar2023data} \\ \hline
Sparse regression techniques & Modeling power converters in microgrids & Constructs precise models for control strategies and stability assessments without continual system perturbations. & \cite{hosseinipour2023sparse,hosseinipour2023data} \\ \hline
SR with sparse dictionary learning & Refining power system load models & Enhances precision and utility of load models by amalgamating diverse mathematical functions. & \cite{10194488} \\ \hline
SINDy with model predictive control & Voltage restoration and power allocation & Reduces voltage recovery time and facilitates precise power distribution among resources. & \cite{tran2023sparse,shadaei2024data} \\ \hline
SINDy & Modeling and controlling distributed energy resources & Effective in safeguarding stability and optimizing performance across diverse grid environments. & \cite{khazaei2023data,khazaei2023data1} \\ \hline
SINDy & Managing modern power grids with renewable energy & Pinpoints crucial features and delineates system-level nonlinearity via an innovative index. & \cite{en17030711} \\ \hline
SINDy & Estimating power grid parameters & Excels in accurately estimating power grid parameters, particularly in systems with slow dynamics. & \cite{su14042051,9862464} \\ \hline
SINDy & Identifying sources of forced oscillations in power systems & Pinpoints sources of oscillations without relying on a pre-existing model, suitable for real-time applications. & \cite{9822991} \\ \hline
SINDy & Enhancing wind farm models & Constructs a comprehensive library of potential dynamics tailored to wind farms, refining simulations. & \cite{wang2023error} \\ \hline 
\hline
\end{tabular}
\end{center}
\end{table*}

This method also demonstrates promising and inspiring results in identifying the origins of forced oscillations emanating from power plants. As detailed in \cite{9822991}, it showcases an ability to pinpoint the sources of such oscillations within power systems without relying on a pre-existing model. This capability is underscored by its adeptness in accurately identifying sources across diverse scenarios, rendering it apt for real-time applications. Notably, its low computational overhead and minimal tuning requisites further enhance its suitability for practical deployment in real-world settings, highlighting its potential as a valuable tool in the detection and mitigation of forced oscillations within power grids.

Furthermore, the authors strive to enhance the precision of models utilized for wind farms by devising a data-driven approach aimed at mitigating errors encountered during dynamic transitions, as elucidated in \cite{wang2023error}. This entails constructing a comprehensive library of potential dynamics tailored to wind farms and harnessing regression algorithms to refine simulations. Through this endeavor, the study demonstrates notable advancements in accurately simulating wind power plants, particularly under challenging conditions such as low voltage ride-through scenarios.

The extensive application of SINDy within power systems is  illustrated in Table \ref{tab:sindy}. This method has been employed across a wide array of contexts, ranging from the modeling of synchronous generators and distributed energy resources to the enhancement of microgrid stability and reliability. The comprehensive scope of this table underscores the substantial commitment to utilizing SINDy to tackle diverse challenges inherent in this field. The versatility and robust performance of SINDy in addressing complex, nonlinear issues within power systems are clear from these varied applications. Moving forward, future research directions could concentrate on promising areas such as state estimation, sensorless control of motors, and motor design. These applications capitalize on SINDy's capacity to learn parsimonious models, handle large datasets effectively, and optimize system configurations and control strategies. Section \ref{sec:5B2} will provide a more detailed examination of these potential applications, underscoring the ongoing exploration of SINDy's capabilities within the field of power systems.

\subsection{Deep SR}

This section explores the practical implementation of deep SR, which involves a novel hybrid approach integrating genetic programming with neural network, reinforcement learning, and the utilization of AI-Feynman to unveil mathematical equations governing power flow. These cutting-edge methodologies represent the forefront of research, dedicating considerable time and resources to apply them to power systems. This endeavor aims to thoroughly examine the strengths and limitations of these methods and facilitate a comprehensive comparison with conventional techniques such as genetic programming and SINDy.

The fusion of genetic programming with neural network has proven to be a practical and effective method for forecasting. Mohamed Trabelsi et al. \cite{en15239008} introduce an innovative approach to predicting solar power output. This approach combines SR with deep multi-layer perceptron models, resulting in a hybrid model that outperforms standalone SR and multi-layer perceptron models in terms of forecasting accuracy and robustness, particularly in predicting photovoltaic power output. Importantly, this methodology holds promise for applications in economic dispatch and unit commitment, offering broad utility across diverse domains.

Furthermore, this hybrid approach extends to electricity load prediction as well. Dimoulkas et al. \cite{dimoulkas2018hybrid} propose a novel technique that integrates SR for feature selection and neural network for precise load forecasts. This hybrid model demonstrates effectiveness in predicting electricity load, showcasing competitive performance in real-world competitions. Validation of this method involves comparisons with standalone implementations of genetic programming and neural network, reinforcing its efficacy.

\begin{table*}[h!]
\begin{center}
\footnotesize
\captionsetup{skip=-0.5pt}
\caption{\raggedright Applications of deep SR in power systems, highlighting their key features and references.}
\label{tab:dsr}
\begin{tabular}{>{\raggedright\arraybackslash}p{0.2\textwidth}>{\raggedright\arraybackslash}p{0.25\textwidth}>{\raggedright\arraybackslash}p{0.35\textwidth}>{\centering\arraybackslash}p{0.09\textwidth}}
\hline \hline
\textbf{Methodology} & \textbf{Application} & \textbf{Key Features} & \textbf{References} \\ \hline
Genetic programming with neural networks & Solar power output prediction & Combines SR with deep multi-layer perceptron models for accurate forecasting. & \cite{en15239008} \\ \hline
Genetic programming with neural networks & Electricity load prediction & Integrates SR for feature selection and neural networks for precise load forecasts, showcasing competitive performance. & \cite{dimoulkas2018hybrid} \\ \hline
Genetic programming with neural networks & Wind power prediction & Enhances prediction accuracy while mitigating errors from atmospheric fluctuations using various neural networks types. & \cite{ZAMEER2017361} \\ \hline
Straight-line programming with genetic programming & Energy consumption modeling & Employs ant colony optimization and straight-line programming for more accurate modeling compared to traditional genetic programming methods. & \cite{RUEDA202023} \\ \hline
Deep learning techniques & Proportional-integral current controller for surface-mounted permanent magnet synchronous motors & Introduces a novel current controller using deep learning techniques, specifically an recurrent neural network, for improved motor control. & \cite{usama2022data} \\ \hline
AI-Feynman & Power flow mathematical expression & Modifies AI-Feynman to determine power flow mathematical expressions in a distribution network. & \cite{anonymous2023physicsbased} \\ \hline
Reinforcement learning & Power converter design & Utilizes the autonomous topology generator to automatically generate efficient circuit topologies for power converters. & \cite{10215513} \\ \hline
Reinforcement learning & Mathematical expression discovery & Facilitates user-centric discovery of mathematical expressions with feedback for SR tasks and power system data exploration. & \cite{ijcai2022p849} \\ \hline \hline
\end{tabular}
\end{center}
\end{table*}

Expanding on the domain of forecasting, in \cite{ZAMEER2017361}, this approach is employed for predicting wind power through a fusion of neural network and genetic programming. The objective here is to enhance prediction accuracy while mitigating errors resulting from atmospheric fluctuations. The study explores the utilization of five different types of neural networks for data organization, with the most suitable network chosen for input into genetic programming to formulate a mathematical expression describing the dataset. Testing conducted on data from five wind farms in Europe yields promising results in terms of error reduction measures.

In a manner akin to forecasting, this approach extends its utility to system modeling as well. R. Rueda et al. \cite{RUEDA202023} harness this integrated methodology to optimize real energy consumption modeling, employing ant colony optimization alongside straight-line programming. Through a comparative analysis against traditional genetic programming methods, the combined approach of straight-line programming and genetic programming demonstrates more accurate performance in modeling energy consumption.

Moreover, this methodology finds application in determining the mathematical expression governing the proportional-integral current controller for surface-mounted permanent magnet synchronous motors \cite{usama2022data}. This study introduces a novel current controller tailored specifically for this motor type, integrating deep learning techniques to establish an optimal control model. Departing from convolutional neural networks, a recurrent neural network is employed for motor control. The primary aim is to enhance performance compared to traditional control methods by utilizing a more adaptable and precise model.

Among the less conspicuous techniques within the realm of deep SR lie AI-Feynman and reinforcement learning. To the author's knowledge, only one paper has employed AI-Feynman in the context of power systems to determine the mathematical expression governing power flow. In \cite{anonymous2023physicsbased}, AI-Feynman underwent modifications, such as omitting the polynomial fit step and substituting the physics-informed neural network with a conventional neural network. These adaptations were applied to the 34-node distribution network, marking a notable endeavor within the field. 

Reinforcement learning has been harnessed in two distinct applications within the domains of power systems. In \cite{10215513}, a study explores the utilization of artificial intelligence to streamline the design process of power converters. This research introduces the autonomous topology generator, a novel approach employing reinforcement learning to automatically generate circuit topologies tailored to user specifications and preferences. By navigating within component constraints, autonomous topology generator aims to produce highly efficient designs, potentially uncovering novel and optimized configurations. Validation of this methodology involves its application to various converter types such as buck, boost, and buck-boost converters. 

In \cite{ijcai2022p849}, reinforcement learning is leveraged to facilitate a user-centric approach in discovering mathematical expressions suited to individual preferences. This method introduces a platform where users can guide the expression-finding process by providing feedback through various means, including categorization, comparison, and suggestion of improvements. This interactive platform is designed for SR tasks and power system data exploration, accessible via a user-friendly web interface, offering a practical tool for researchers and practitioners alike.

Deep SR methodologies hold great promise for applications in power systems, offering insights into their capabilities and constraints. A comprehensive overview of various methodologies and their applications in this field is provided in Table \ref{tab:dsr}, highlighting significant efforts aimed at harnessing deep SR for tasks such as predicting solar power output, electricity load forecasting, and wind power estimation. These applications underscore the methodology's adaptability and robust performance in addressing the intricate, nonlinear challenges inherent in power systems. Looking ahead, future research could focus on specific promising areas including optimal control, fault detection and diagnosis, and dynamic state estimation. These applications capitalize on deep SR's ability to generate interpretable models, handle large datasets effectively, and optimize control strategies. Section \ref{sec:5C2} will delve deeper into these potential applications, emphasizing how deep SR excels in interpretability, flexibility, and automated discovery of mathematical relationships within power systems.

\section{Comparative Study of SR Methods}
\label{sec:4}

In this section, three detailed case studies are presented to enable a comprehensive comparison of SR methodologies. The objective is to rigorously examine the advantages and limitations inherent in each approach. Subsequently, research gaps and future directions related to each method that warrant further investigation are identified in section \ref{sec:5}. In order to identify the underlying dynamics in the data generated for each case study, we utilized the following Python packages for each method: the SINDy method \cite{desilva2020pysindy, Kaptanoglu2022}, and the deep SR technique \cite{meurer2017sympy, paszke2017autodiff}, each tailored to the respective analysis. Notably, in the case of deep SR, we employed a combination of \cite{meurer2017sympy} and \cite{paszke2017autodiff} for comparative analysis. To implement the ARGOS methodology, we utilized the ARGOS package within the R programming environment \cite{egan2024automatically}. It is important to note that for each case study, after generating the dataset, we split it into 80\% for training and 20\% for testing. The system dynamics were identified using the training dataset and subsequently validated on the test dataset. The results presented here correspond to the testing stage. All simulations were executed on a workstation running Windows 11 Pro 64-bit. The workstation was equipped with a 13th Gen Intel(R) Core(TM) i9-13900H processor running at 2.60 GHz and 32 GB of memory.

\subsection{Single Machine Infinite Bus System}
\begin{figure}[!ht]
\centering
\begin{tikzpicture}[line cap=round,line join=round,>=triangle 45,x=1cm,y=1cm,scale=1.5]
\clip(8.3,3.55) rectangle (14.2,4.7);
\draw [line width=0.4pt] (8.8,4) circle (0.4cm);
\draw [line width=0.4pt] (9.5,4.6)-- (9.5,3.6);
\draw [line width=0.4pt] (11.6,4.6)-- (11.6,3.6);
\draw [line width=1.2pt] (14,4.6)-- (14,3.6);
\draw [line width=0.4pt] (13.2,4)-- (14,4);
\draw [line width=0.4pt] (12.4,4.1)-- (12.4,3.9);
\draw [line width=0.4pt] (12.4,3.9)-- (13.2,3.9);
\draw [line width=0.4pt] (13.2,3.9)-- (13.2,4.1);
\draw [line width=0.4pt] (13.2,4.1)-- (12.4,4.1);
\draw (9.17,4.62) node[anchor=north west] {$\overline{E}$};
\draw (11.17,4.62) node[anchor=north west] {$\overline{V}_{1}$};
\draw (13.57,4.62) node[anchor=north west] {$\overline{V}_{2}$};
\draw (9.95,4.48) node[anchor=north west] {$jX^{\prime}_{d}$};
\draw (12.33,4.46) node[anchor=north west] {$R+jX$};
\draw (10.7,4.00) node[anchor=north west] {$P+jQ$};
\draw (8.5,4.176831819712612) node[anchor=north west] {Gen.};
\draw [line width=1.2pt] (14.1,4.48)-- (14,4.38);
\draw [line width=1.2pt] (14.1,4.6)-- (14,4.5);
\draw [line width=1.2pt] (14.1,4.54)-- (14,4.44);
\draw [line width=0.4pt] (9.8,4.1)-- (9.8,3.9);
\draw [line width=0.4pt] (9.8,3.9)-- (10.6,3.9);
\draw [line width=0.4pt] (10.6,3.9)-- (10.6,4.1);
\draw [line width=0.4pt] (10.6,4.1)-- (9.8,4.1);
\draw [line width=0.4pt] (9.2,4)-- (9.8,4);
\draw [line width=0.4pt] (10.6,4)-- (12.4,4);
\draw [line width=0.8pt] (11.26,4)-- (11.54,4);
\begin{scriptsize}
\draw [fill=black,shift={(11.54,4)},rotate=270] (0,0) ++(0 pt,1.5pt) -- ++(1.299038105676658pt,-2.25pt)--++(-2.598076211353316pt,0 pt) -- ++(1.299038105676658pt,2.25pt);
\end{scriptsize}
\end{tikzpicture}
\vspace{-.6cm}
\caption{One-line diagram of a single-machine infinite bus system with parameters in p.u.: R=0.05, X=0.3, $\overline{V_1}$=1.05, $\overline{V_2}$=1.0, P=0.8, $X'_d$=0.2 \cite{9308946}.}
\label{fig:onelinediagram}
\end{figure}
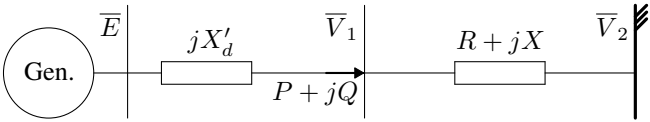

The single machine infinite bus system serves as a fundamental model in power system analysis, illustrating the dynamic behavior and stability of a single generator connected to an infinite bus, which represents a large, unchanging power grid via a transmission line. The concise single-line diagram, illustrated in Fig. \ref{fig:onelinediagram}, encompasses all the attributes of the system.

Initially, the single machine infinite bus system is at equilibrium, and at t=1 second, a discrete disturbance with a magnitude of 0.17 per-unit (p.u.) is introduced to analyze the transient response. Data is collected from this disturbed system to allow a thorough evaluation of the system's dynamic behavior. The dataset is generated in a Python environment, based on the power flow and swing equations (Eq. \eqref{power flow} and Eq. \eqref{swing}), which provide the variation of the generator's speed at bus 1 \cite{9308946}.

\begin{equation}
    P_{12} = G_{12}V_1^2 - G_{12}V_1V_2cos(\delta) - B_{12}V_1V_2sin(\delta)
\label{power flow}
\end{equation}
\begin{equation}
    \frac{\partial \delta}{\partial t} = \omega, \qquad \qquad \frac{\partial \omega}{\partial t} = \frac{1}{M}(P_m - P_e - D\omega)
\label{swing}
\end{equation}
\begin{equation*}
    G = \frac{R}{R^2 + X^2} \qquad \qquad B = \frac{-X}{R^2 + X^2}
\end{equation*}

Let $\delta$ represent the rotor angle, and 
$\omega$ denote the rotor speed. The term $P_{12}$ indicates the power flow between nodes 1 and 2, while $P_{m}$ refers to the mechanical power, and $P_{e}$ signifies the electrical power. The variables $V_{1}$ and $V_{2}$ correspond to the magnitudes of voltage at buses 1 and 2, respectively. Additionally, $G$ and 
$B$ represent conductance and susceptance, while $D$ denotes the damping coefficient and $M$ indicates the inertia constant of the machine. Once the dataset is generated from this disturbed system, various SR methodologies are applied to analyze the changes in rotor speed throughout the simulation period. By comparing these methods, we gain insights into their relative strengths and weaknesses in modeling the system dynamics efficiently and accurately. The performance of the SR methods — SINDy, ARGOS, and deep SR — in terms of mean squared error, $R^2$ score, and total computational time is summarized in Table \ref{tab:performance}.

\begin{table}[ht!]
\centering
\caption{Performance comparison of SINDy, ARGOS, deep SR to the generated dataset for single machine infinite bus system based on mean square error, $R^2$ score, and total time elapsed.}
\begin{tabular}{|>{\centering\arraybackslash}m{2cm}|>{\centering\arraybackslash}m{1.5cm}|>{\centering\arraybackslash}m{1.5cm}|>{\centering\arraybackslash}m{1.5cm}|}
\hline
\textbf{Method} & \textbf{Mean Square Error} & \textbf{\boldsymbol{$R^2$} Score} & \textbf{Total Time Elapsed (seconds)} \\ 
\hline
%\textbf{Genetic programming} & 7.4e-06 & -0.001 & 4.58 \\ 
%\hline
\textbf{SINDy} & 1.7e-10 & 0.999 & 0.02 \\ 
\hline
\textbf{ARGOS} & 4.5e-9 & 0.999 & 1.9 \\ 
\hline
\textbf{Deep SR} & 2.6e-10 & 0.999 & 10.51 \\ 
\hline
\end{tabular}
\label{tab:performance}
\end{table} 

%To further illustrate the effectiveness of each method, Fig. \ref{fig:performance} presents visual comparisons of their accuracy in capturing the system's dynamics and the corresponding absolute error for each method. 
%Genetic programming, as expected, struggles to follow variations in the dataset due to its inherent bias towards algebraic expressions. Since the data is derived from ordinary differential equations, genetic programming's limitations are evident.
% Academic Discussion of Speed Variation and Error Analysis
%The Fig. \ref{fig:performance} presents the variation of speed and corresponding absolute errors over time for a system subjected to a disturbance at $t = 1$~s. 
%The upper subplot compares the true variation of speed with predictions from all three methods: SINDy, ARGOS, and Deep SR. All methods exhibit excellent agreement with the true system dynamics, accurately capturing the transient oscillations and their decay. This demonstrates their effectiveness in modeling nonlinear dynamics in power systems. However, minor discrepancies are visible during the transient phase, with SINDy and Deep SR providing slightly better alignment than ARGOS. 

\begin{figure}[htbp]
    \centering
        \centering
        \includegraphics[width=0.5\textwidth]{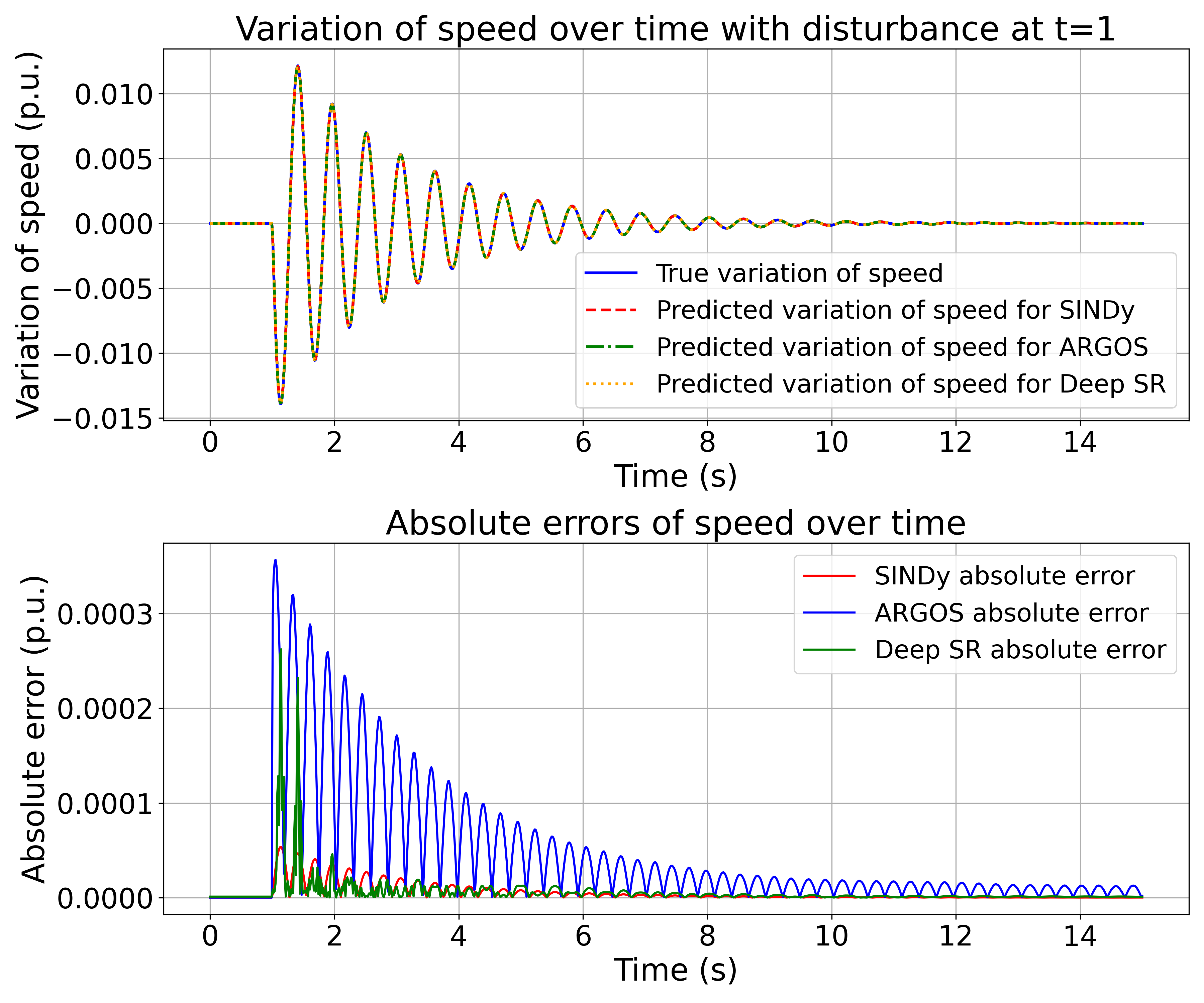}
    \caption{Performance comparison of SR methods applying to the generated dataset for single machine infinite bus system. %(a) Genetic programming; 
    %(a) SINDy; (b) ARGOS; (c) Deep SR.
    }
\label{fig:performance}
\end{figure}

To further illustrate the effectiveness of each method, Fig. \ref{fig:performance} compares their accuracy in capturing system dynamics and presents the corresponding absolute errors.
%The upper subplot depicts the true speed variation alongside predictions from SINDy, ARGOS, and deep SR. 
The prediction trajectories of all methods align closely with the true system dynamics, effectively modeling transient oscillations and their decay. Minor discrepancies are observed during the transient phase, where SINDy and deep SR demonstrate better alignment than ARGOS. During the critical transient phase, SINDy and deep SR exhibit lower error magnitudes compared to ARGOS. ARGOS, though initially less accurate, achieves significant improvement in the steady-state period. Over time, all methods converge to minimal steady-state errors. These findings underscore the suitability of SINDy and deep SR for applications requiring high accuracy in transient dynamics, while ARGOS may be more apt for steady-state analyses.

%The lower subplot highlights absolute speed prediction errors. 

Despite their comparable accuracy, computation time varies significantly. Deep SR, taking approximately 10.5 seconds, highlights the inherent complexity of neural networks. In contrast, SINDy achieves the lowest error with a response time of just 0.02 seconds, making it highly efficient for real-time applications. This trade-off between accuracy and efficiency emphasizes SINDy's potential for tasks like real-time system identification and control in power networks.

%The lower subplot highlights the absolute errors in speed predictions. SINDy and Deep SR show smaller error magnitudes compared to ARGOS, especially during the critical transient phase following the disturbance. While ARGOS exhibits larger initial errors, its performance improves significantly in the steady-state period. Overall, all methods achieve minimal errors over time, with steady-state errors converging to negligible levels. These results indicate the potential of SINDy and Deep SR for applications requiring high accuracy in transient dynamics, while ARGOS may be more suited for steady-state analyses. This comparison underscores the reliability and robustness of these approaches for system identification in power networks.

%Although SINDy, ARGOS, and deep SR successfully capture the system dynamics, deep SR takes significantly longer (approximately 10.5 seconds) due to the inherent complexity of neural networks. This extended computation time demonstrates a trade-off between accuracy and efficiency. SINDy, with its notably faster response time (0.02 seconds), provides the lowest error and highest computational efficiency, making it ideal for real-time applications.

%Based on this comparison, SINDy emerges as the best candidate for fast and accurate model identification. Consequently, in subsequent case studies, we focus primarily on SINDy and deep SR, as genetic programming proves inadequate for this particular task of capturing system dynamics.

\subsection{Grid-following Inverter}

Grid-following inverters are essential components in modern renewable energy systems, enabling the integration of distributed energy resources into the power grid. Understanding their dynamic behavior is crucial for optimizing performance and ensuring stability, especially as power grids evolve to accommodate increasing renewable energy penetration. In this work, we aim to model and capture the dynamics of a grid-following inverter using the SINDy, ARGOS, and deep SR methods to compare their effectiveness.

The dynamics of the grid-following inverter are represented in the 
dq-frame, where \( i_d \) and \( i_q \) denote the direct and quadrature axis currents, respectively. The dq-frame is commonly used for analysis in power systems as it simplifies the representation of AC signals by transforming them into a rotating reference frame. The governing equations for the system are derived from the voltage and current relationships in the per-unit system, with the per-unit values for inductance \( L_{\text{pu}} \)  = 0.0189 and resistance \(R_{\text{pu}} = 1.89\) used for the analysis. With initial conditions set to zero, the system is simulated over a time span of 0 to 0.1 seconds to capture the transient dynamics. The equations governing the system’s behavior are given by:
\begin{align}
\frac{di_d}{dt} & = \frac{v_d - {R_{pu}} \cdot i_d + \omega {L_{pu}} \cdot i_q}{L_{pu}} \\
\frac{di_q}{dt} & = \frac{v_q - {R_{pu}} \cdot i_q - \omega {L_{pu}} \cdot i_d}{L_{pu}}
\end{align}

Where $v_d$ and $v_q$ are direct and quadrature axis voltages. $R_{pu}$, $L_{pu}$, and $\omega$ are resistance, inductance, and grid angular frequency, respectively. After generating the dataset based on this model, we applied SINDy, ARGOS, and deep SR methods to capture the system dynamics. The effectiveness of these methods was evaluated by comparing their outputs against the simulated inverter behavior using the mean squared error and \( R^2 \) score metrics.  

\begin{table}[htbp]
\centering
\caption{Performance comparison of SINDy, ARGOS and deep SR methods for the grid-following inverter system based on mean squared error, $R^2$ score, and total computation time.}
\begin{tabular}{|>{\centering\arraybackslash}m{1cm}|>{\centering\arraybackslash}m{1cm}|>{\centering\arraybackslash}m{1cm}|>{\centering\arraybackslash}m{1cm}|>{\centering\arraybackslash}m{1cm}|>{\centering\arraybackslash}m{1.25cm}|}
\hline
\textbf{Method} & \textbf{Mean Square Error (\( i_d \))} & \textbf{Mean Square Error (\( i_q \))} & \textbf{\boldsymbol{$R^2$} Score (\( i_d \))} & \textbf{\boldsymbol{$R^2$} Score (\( i_q \))} & \textbf{Total Time Elapsed (seconds)} \\ 
\hline
\textbf{SINDy} & 1.9e-09 & 1.4e-09 & 0.999 & 0.999 & 0.11 \\ 
\hline
\textbf{ARGOS} & 8.3e-09 & 1.04e-07 & 0.999 & 0.999 & 1.88 \\ 
\hline
\textbf{Deep SR} & 9e-08 & 6.7e-08 & 0.999 & 0.999 & 10.15 \\ 
\hline
\end{tabular}
\label{tab:performance1}
\end{table}

\begin{figure}[htbp]
\centering
\includegraphics[width=0.5\textwidth]{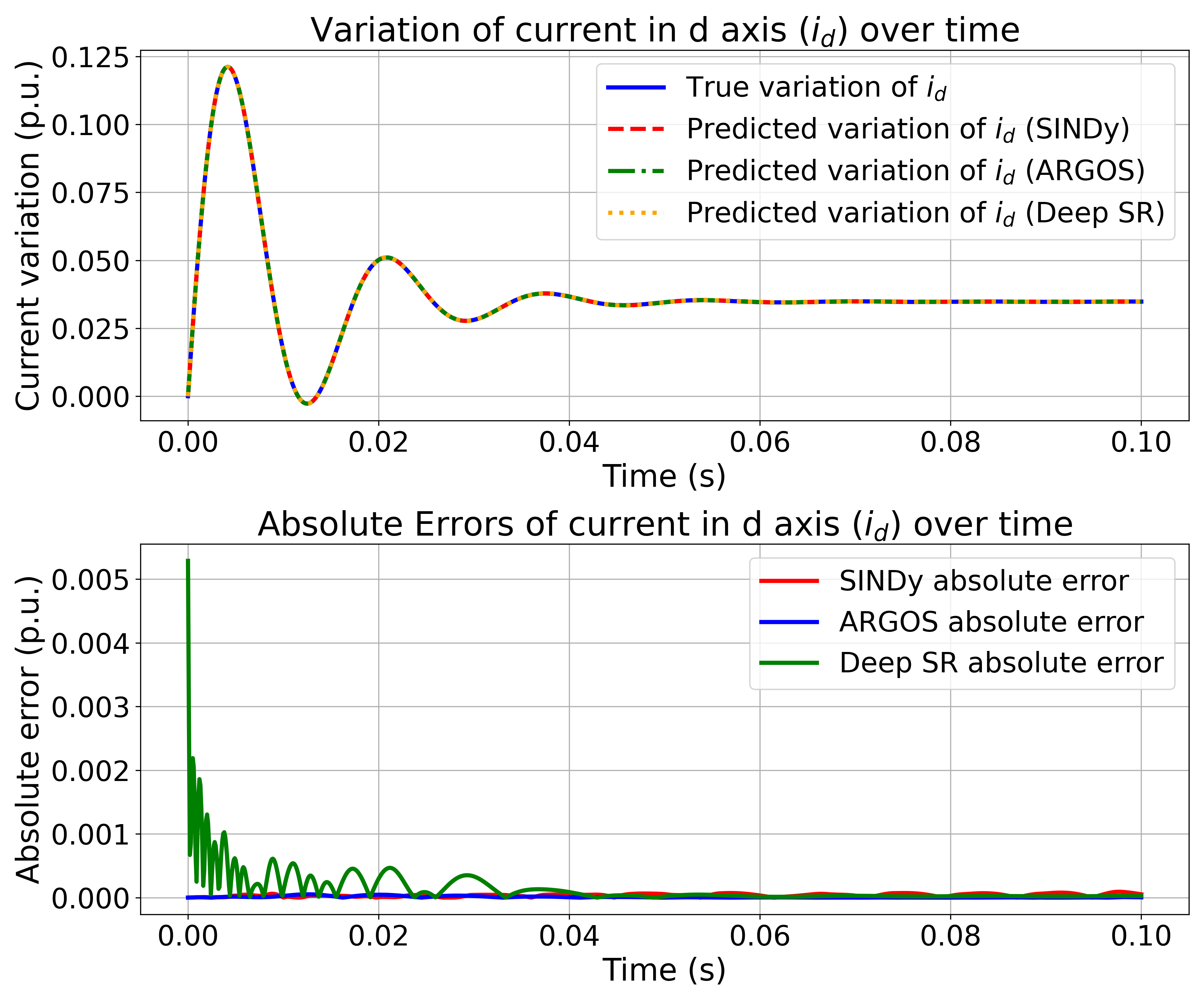} 
    \caption{Performance comparison of SR methods applied to the generated dataset for a grid-following inverter, focusing on the current in the d-axis ($i_d$). %(a) SINDy; (b) ARGOS; (c) Deep SR.
    }
\label{fig:performance_id}
\end{figure}

\begin{figure}[htbp]
\centering
\includegraphics[width=0.5\textwidth]{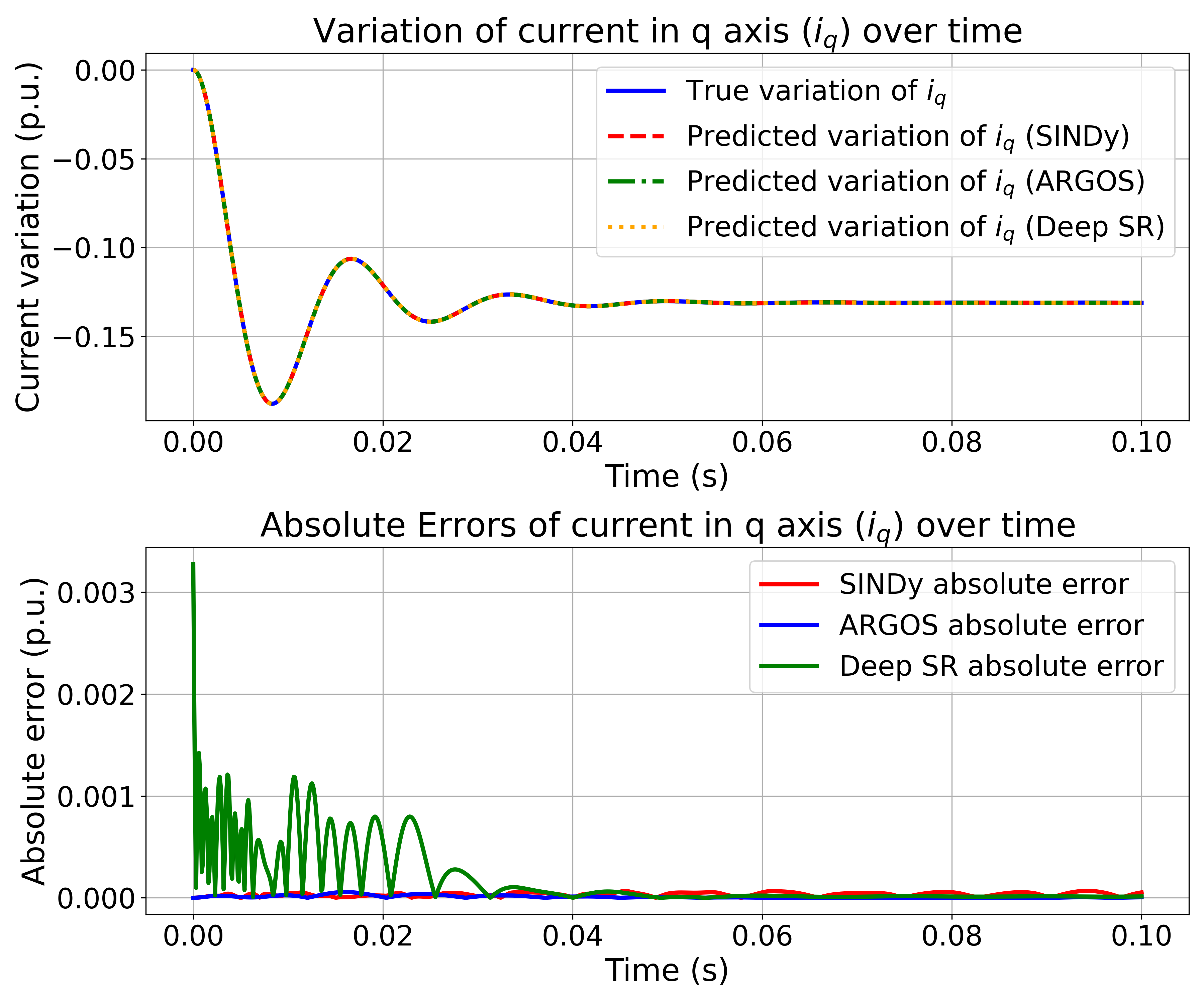} 
    \caption{Performance comparison of SR methods applied to the generated dataset for a grid-following inverter, focusing on the current in the q-axis ($i_q$). %(a) SINDy; (b) ARGOS; (c) Deep SR.
    }
\label{fig:performance_iq}
\end{figure}

A detailed comparison of the accuracy and computational efficiency of SINDy, ARGOS, and deep SR in capturing the inverter dynamics and the corresponding absolute error for each method is provided in Table \ref{tab:performance1}, Fig. \ref{fig:performance_id}, and Fig. \ref{fig:performance_iq}. All methods demonstrate high accuracy, with $R^2$ scores approaching 1, and effectively track the system's dynamic trajectories. Although all methods exhibit satisfactory performance in terms of absolute errors for both state variables (\( i_d \), and \( i_q \)), deep SR shows higher absolute errors compared to the other two methods during the system's initial phase. However, it successfully minimizes absolute errors over time.  %The true and predicted variations of \(i_d\) align closely for all methods. This indicates robust modeling of d-axis dynamics. Similar to \(i_q\), the transient response highlights the models' capability to manage rapid system changes. 
%The absolute error plot shows SINDy achieving the smallest error, with ARGOS and Deep SR performing slightly worse during the transient phase. Despite this, all models demonstrate rapid error decay, achieving high accuracy in the steady-state.
%The top plot shows the true variation of \(i_q\) and its predictions using SINDy, ARGOS, and Deep Symbolic Regression (Deep SR). 
%All three methods closely match the true dynamics, particularly during steady-state conditions. In the transient region (initial 0.01 seconds), the models effectively capture rapid changes, with SINDy demonstrating the highest precision.
%The bottom plot depicts the absolute error. SINDy consistently exhibits the smallest error magnitude, while ARGOS and Deep SR show slightly higher peaks during transients. The errors stabilize to near-zero values as the system approaches steady-state, demonstrating the methods' accuracy over time.
It is important to note that deep SR requires significantly more computational time compared to SINDy, and ARGOS due to the complexity of the neural network-based approach. This highlights the trade-off between accuracy and computational efficiency when using data-driven modeling techniques for modeling.

\subsection{Grid-forming Inverter}

Grid-forming inverters play a critical role in modern power systems by providing voltage and frequency stability through the emulation of synchronous generator characteristics. These inverters have become particularly important in renewable energy systems due to their ability to stabilize voltage and frequency in weak grids or during disturbances. Acting as voltage sources, the dynamics of these inverters are essential for ensuring reliable grid operation, especially under sudden disturbances. 

A dynamic model of a grid-forming inverter operating in a per-unit system is presented in Eq. \eqref{vd}, and Eq. \eqref{vq}. The model captures the behavior of the direct (\( v_d \)) and quadrature (\( v_q \)) voltage components through differential equations. The system is governed by the following equations:

Adjusted droop control:
\begin{equation}
    i_d = \frac{V_{ref} - v_d}{R_{pu}}, \quad
    i_q = -\frac{v_q}{R_{pu}}
\end{equation}

Voltage dynamics:
\begin{equation}
    \frac{dv_d}{dt} = \frac{\omega L_{pu} i_q - R_{pu} i_d}{L_{pu}}
    \label{vd}
\end{equation}
   
\begin{equation}
    \frac{dv_q}{dt} = \frac{-\omega L_{pu} i_d - R_{pu} i_q}{L_{pu}}
    \label{vq}
\end{equation}

Where $i_d$ and $i_q$ are direct and quadrature axis currents.
$R_{pu}$, $L_{pu}$, $\omega$, and $V_{ref}$ are resistance, inductance, grid angular frequency, and reference voltage, respectively. The dynamic behavior of the system was simulated over a time span of 0 to 0.1 seconds, with values of 0.0189 for ${L_{pu}}$ and 1.89 for ${R_{pu}}$ and initial conditions set to zero. The dataset generated from this simulation was used to identify the system's dynamics using SINDy, ARGOS, and deep SR methods. 

\begin{figure}[htbp]
    \centering
        \includegraphics[width=0.5\textwidth]{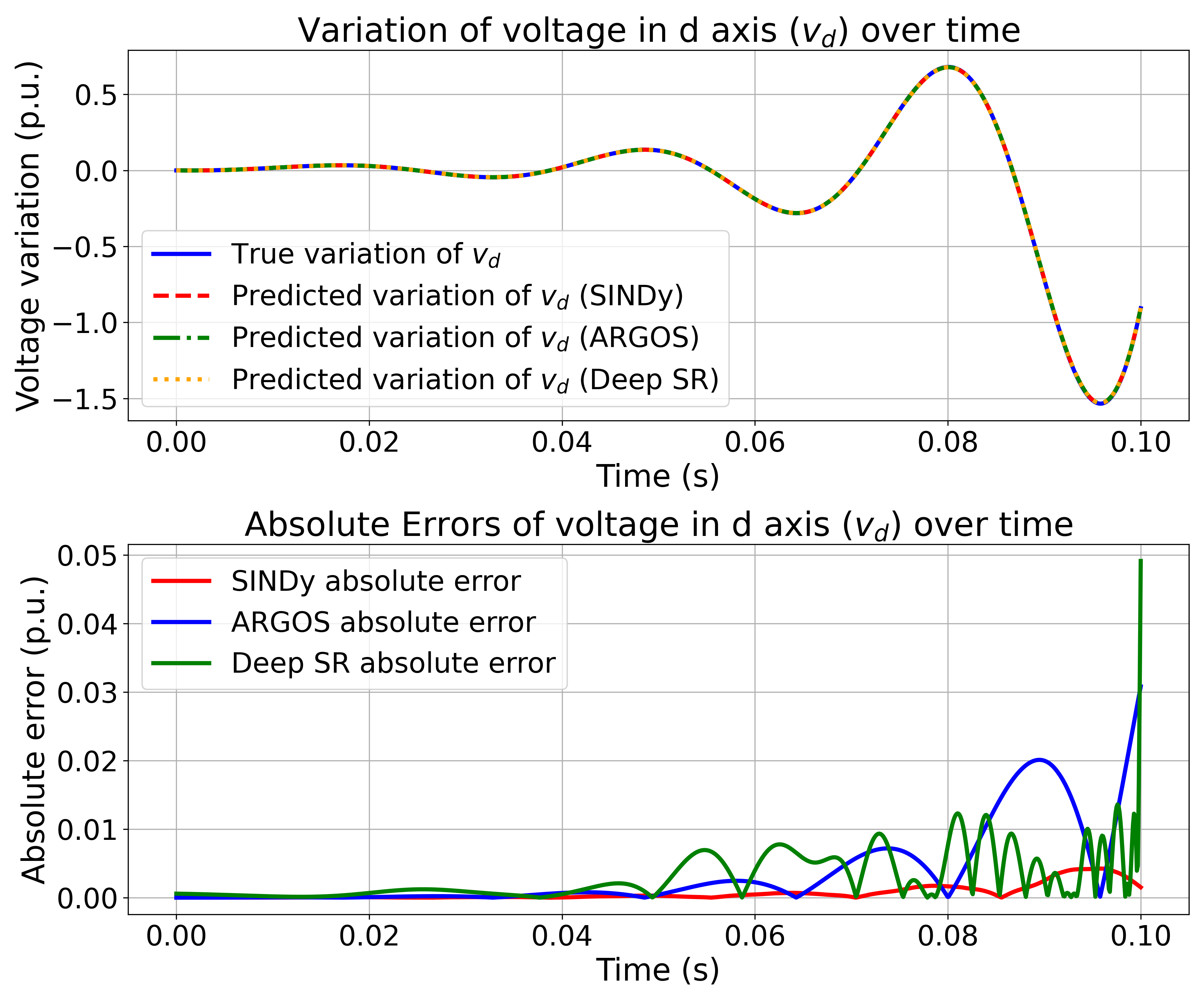} 
    \caption{Performance comparison of SR methods applied to the generated dataset for a grid-forming inverter, focusing on the voltage in the d-axis ($v_d$). %(a) SINDy; (b) ARGOS; (c) Deep SR.
    }
\label{fig:performance_vd}
\end{figure}

\begin{figure}[htbp]
    \centering
        \includegraphics[width=0.5\textwidth]{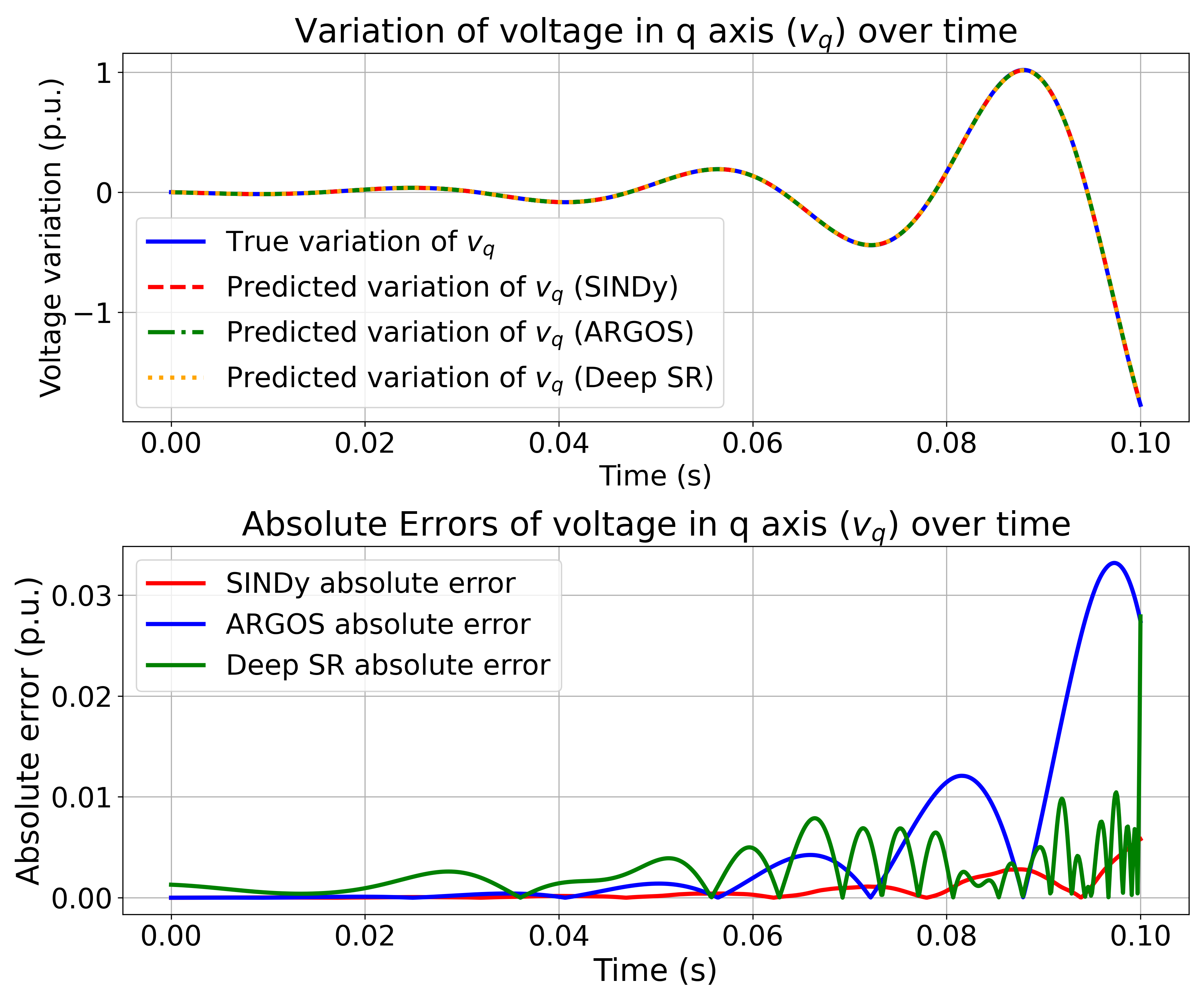} 
    \caption{Performance comparison of SR methods applied to the generated dataset for a grid-forming inverter, focusing on the voltage in the q-axis ($v_q$). %(a) SINDy; (b) ARGOS; (c) Deep SR.
    }
\label{fig:performance_vq}
\end{figure}

To evaluate the model's performance, metrics from two other case studies—mean squared error and $R^2$ score—were used. The accuracy, and computational efficiency related to all methods in capturing the underlying dynamics is illustrated in Table \ref{tab:performance2}, Fig. \ref{fig:performance_vd}, and Fig. \ref{fig:performance_vq}, along with the corresponding absolute error for each method, with deep SR requiring more computational resources than SINDy. While all methods effectively capture the system's dynamics, both deep SR and ARGOS exhibit suboptimal performance in terms of absolute error over time. In contrast, SINDy demonstrates significantly greater accuracy and provides a more satisfactory performance in this case study. Analysis of the metrics, trajectories, and corresponding absolute errors for each method clearly demonstrates that SINDy provides a more accurate and faster response compared to the other methods in capturing the system's dynamics, whereas ARGOS produces more concise mathematical expressions due to its trimming capability.

\begin{table}[htbp]
\centering
\caption{Performance comparison of SINDy, ARGOS and deep SR methods for the grid-forming inverter system based on mean squared error, $R^2$ score, and total computation time.}
\begin{tabular}{|>{\centering\arraybackslash}m{1cm}|>{\centering\arraybackslash}m{1cm}|>{\centering\arraybackslash}m{1cm}|>{\centering\arraybackslash}m{1cm}|>{\centering\arraybackslash}m{1cm}|>{\centering\arraybackslash}m{1.25cm}|}
\hline
\textbf{Method} & \textbf{Mean Square Error (\( v_d \))} & \textbf{Mean Square Error (\( v_q \))} & \textbf{\boldsymbol{$R^2$} Score (\( v_d \))} & \textbf{\boldsymbol{$R^2$} Score (\( v_q \))} & \textbf{Total Time Elapsed (seconds)} \\ 
\hline
\textbf{SINDy} & 1.6e-06 & 1.5e-06 & 0.999 & 0.999 & 0.36 \\ 
\hline
\textbf{ARGOS} & 4.8e-05 & 8.5e-05 & 0.999 & 0.999 & 1.9 \\ 
\hline
\textbf{Deep SR} & 2.1e-05 & 1.4e-05 & 0.999 & 0.999 & 9.53 \\ 
\hline
\end{tabular}
\label{tab:performance2}
\end{table}

%It is important to recognize that these basic examples do not allow for a comprehensive and equitable comparison of all methodologies. The datasets used in these case studies take into account the changes that occur after applying a disturbance, which are formulated using ordinary differential equations. Therefore, it is anticipated that SINDy will outperform other techniques, as it is specifically designed for dynamical systems involving ordinary and partial differential equations.

A course for future inquiry within the domain of power systems will be represented, offering a comprehensive road-map for further investigation. Section \ref{sec:3} extensively examines the practical applications of each method, leading to a comprehensive understanding of the specific domains wherein these methods find utility. It is imperative to initiate this exploration by delineating the advantages and disadvantages of each method. Through such an analytical framework, research gaps and potential applications intrinsic to each method can be identified, elucidating their distinct characteristics and highlighting areas conducive to further investigation and advancement.

\section{Critical Analysis of SR Methods}
\label{sec:5}

This section aims to provide a thorough academic analysis to assess the strengths, limitations, research gaps, and potential applications of each method. The goal is to offer researchers a structured framework for their investigations, thereby supporting future exploratory efforts in the field. The discussion begins with an examination of SINDy, proceeds to ARGOS, and concludes with an exploration of deep SR.

\subsection{Sparse Identification of Nonlinear Dynamics (SINDy)}
\subsubsection{Strengths and Limitations}

SINDy possesses its own set of advantages and disadvantages worthy of discussion. This method offers several advantages in modeling. Firstly, it excels in computational efficiency, making it suitable for analyzing large datasets without significant computational overhead. Secondly, SINDy often produces sparse models, which are simpler and more interpretable compared to traditional nonlinear models. This simplicity enhances the understanding of system dynamics, as it automatically selects relevant terms in the model, reducing the need for manual feature engineering. Moreover, being data-driven, this technique can uncover underlying dynamics directly from observational data, eliminating the requirement for prior knowledge of system equations. The identified sparse model terms not only offer insight into the system's behavior but also facilitate its interpretation.

However, SINDy does come with its limitations. One notable drawback is its potential lack of accuracy in modeling highly complex or noisy systems. Additionally, the performance of SINDy models can be sensitive to parameter choices, such as the sparsity parameter and regularization strength, necessitating careful tuning. Furthermore, this approach assumes that the underlying dynamics can be represented by a linear combination of basis functions, which might not always hold true for highly nonlinear systems. Another limitation is its reliance on observed variables, which could lead to the omission of important unobserved variables or interactions, potentially affecting the model's fidelity. Lastly, without proper regularization, SINDy models may be prone to overfitting, especially in scenarios involving high-dimensional data or limited observations. Despite these drawbacks, SINDy remains a valuable tool in modeling and offers significant advantages in uncovering complex dynamical behaviors from observational data. 

\subsubsection{Research Gaps and Potential Applications}

\label{sec:5B2}
SINDy demonstrates notable strengths in SR, particularly its computational efficiency and capacity to derive sparse, interpretable models directly from observational data. These attributes streamline the modeling process and enhance interpretability without requiring extensive manual feature selection. However, inherent limitations include potential inaccuracies in modeling highly complex or noisy systems and sensitivity to parameter settings, necessitating precise tuning for optimal performance. Addressing these challenges demands advancements in derivative estimation under noisy conditions and adaptive strategies for constructing model libraries. Furthermore, enhancing scalability to high-dimensional systems through advanced algorithms and dimensionality reduction techniques is critical for broadening SINDy’s applicability across diverse domains. The principal research gaps are as follows:

\begin{enumerate}
 \item \textit{Handling Noisy Data}: The reliability of the models identified by SINDy heavily depends on accurate derivative estimation, which is particularly difficult when data is noisy. Developing robust methods for derivative estimation that can withstand high levels of noise is crucial. Additionally, improved preprocessing and noise reduction techniques are needed to clean data before applying SINDy, ensuring that the models derived are accurate and reliable.
 \item \textit{Library Construction}: Developing adaptive methods that can construct the library of candidate functions based on the characteristics of the data or preliminary results would enhance the flexibility and applicability of SINDy. Additionally, incorporating non-polynomial functions, such as rational functions or neural network-based functions, into the library could improve the method’s ability to capture more complex dynamics.
 \item \textit{Scalability to High-Dimensional Systems}: SINDy faces challenges when applied to high-dimensional systems due to computational limitations and the complexity of the models. Integrating advanced dimensionality reduction techniques can help simplify these systems before applying SINDy. Additionally, developing more efficient algorithms to handle large-scale problems and extensive libraries of candidate functions is essential for making SINDy more scalable and practical for high-dimensional applications.
\end{enumerate}

The identified research gaps in SINDy highlight critical areas for advancement that align with its promising applications in various domains. Enhancing derivative estimation methods to mitigate noise and developing adaptive strategies for constructing model libraries are essential for improving the accuracy and reliability of models used in fault detection and diagnosis within power systems. Similarly, addressing scalability issues through efficient algorithms and advanced dimensionality reduction techniques is crucial for effectively applying SINDy in tasks such as sensorless control of motors and predictive maintenance. These applications leverage SINDy’s computational efficiency and sparse modeling capabilities to streamline modeling processes and enhance operational reliability. Nonetheless, managing SINDy’s limitations, such as model accuracy in highly complex systems and assumptions about basis functions, remains imperative for maximizing its utility across these diverse applications. The promising applications in this domain are as follows:

\begin{enumerate} 

 \item \textit{Fault Detection and Diagnosis}: SINDy's computational efficiency and ability to uncover underlying dynamics from observational data make it suitable for identifying fault signatures in power system components such as transformers, generators, and transmission lines. This application leverages SINDy's strength in handling large datasets and simplifying model interpretation, which is crucial for early fault detection and maintaining system reliability. 
 \item \textit{State Estimation}: The method's capacity to produce sparse, interpretable models directly from data can be utilized for state estimation in power systems. By identifying governing equations, SINDy helps build accurate state estimation models, addressing its advantage of eliminating the need for prior knowledge of system equations.
 \item \textit{Sensorless Control of Motors}: Applying SINDy to sensorless control of motors, such as permanent magnet synchronous motors, switched reluctance motors, and induction motors, highlights its strength in modeling and control through data-driven methods. This approach benefits from SINDy's ability to produce simple, interpretable models that can facilitate control strategies without relying on extensive sensor data.
 \item \textit{Motor Design}: Optimizing the design of motors using SINDy involves leveraging its data-driven methods for modeling and optimization. The technique's strength in producing sparse models helps in achieving desired performance metrics by simplifying the highly complex design spaces, making it more manageable and interpretable.
 \item \textit{Predictive Maintenance}: SINDy's ability to analyze historical data and identify patterns indicative of impending equipment failure or degradation aligns with its data-driven nature and computational efficiency. This application addresses its advantage in handling large datasets to develop predictive maintenance strategies, thereby minimizing downtime and extending the lifespan of power system components.
 
\end{enumerate}

However, these applications also highlight SINDy's limitations. The potential lack of accuracy in highly complex or noisy systems and sensitivity to parameter choices must be managed through careful tuning and robust validation techniques. Additionally, ensuring that the assumptions about the linear combination of basis functions hold true and addressing the reliance on observed variables are critical for the successful application of SINDy in these areas. Therefore, thorough experimentation and proper regularization are essential to mitigate these limitations so that researchers can explore these promising applications.

\subsection{Automatic regression for governing equations (ARGOS)}
\subsubsection{Strengths and Limitations}
The ARGOS method offers a powerful approach to automatically discovering governing equations for dynamical systems, especially in noisy and complex environments. Its integration of sparse regression with confidence intervals makes it particularly effective for identifying the most relevant parameters while filtering out less significant terms. This ensures that the resulting models are both interpretable and concise, which is essential in many scientific fields, including power systems. Furthermore, ARGOS excels in noisy data scenarios by employing denoising techniques like the Savitzky-Golay filter, allowing it to handle data with high noise levels while maintaining accuracy. The automation of hyperparameter tuning and the use of bootstrapping to quantify uncertainties make ARGOS a robust tool for researchers, significantly reducing the manual effort typically required for model discovery.

Despite its strengths, ARGOS has some notable limitations. One key drawback is its reliance on the assumption that the governing terms exist within the predefined feature matrix. If the relevant features are not included or are poorly represented in the matrix, the method may fail to capture the true system dynamics, limiting its accuracy. Additionally, the method’s computational cost is a concern, especially when bootstrap sampling is involved, as it can be computationally intensive for large datasets or real-time applications. This issue may restrict its applicability in scenarios that require rapid model identification, such as real-time power system monitoring and control. Moreover, while ARGOS performs well in moderately noisy environments, its performance deteriorates significantly when data quality is extremely poor, which could pose challenges in fields where precise measurements are difficult to obtain.

\subsubsection{Research Gaps and Potential Applications}
While the ARGOS method has proven to be effective in automating the discovery of governing equations from data, several research gaps remain:

\begin{enumerate}
    \item \textit{Scalability to Large-Scale Systems}: Although ARGOS has been successful in identifying equations for low- to medium-dimensional systems, its scalability to large-scale, high-dimensional systems, such as complex power grids with numerous dynamic components, has not been fully explored. Further research is needed to evaluate its performance in such settings.

    \item \textit{Real-Time Applications}: ARGOS has primarily been tested on static datasets. The method's ability to handle real-time, non-stationary data, particularly in dynamic environments like power system operation or grid stability monitoring, requires further investigation. Incorporating adaptive learning or online updating mechanisms could enhance ARGOS’s suitability for real-time applications.
    
    \item \textit{Handling Highly Corrupted Data}: Although ARGOS performs well in moderately noisy environments, its robustness against highly corrupted data remains a challenge. Developing more sophisticated noise-handling techniques or integrating methods that better manage low signal-to-noise ratios would improve the reliability of the models generated by ARGOS.
    
    \item \textit{Computational Efficiency}: The current implementation of ARGOS involves bootstrap sampling and repeated regression steps, which can be computationally intensive. Optimizing the algorithm to reduce computational overhead, especially for real-time or large-scale applications, is a critical area for improvement.
    
    \item \textit{Generalization Beyond Known Features}: ARGOS relies on the assumption that the governing terms exist within a predefined feature matrix. Exploring methods to dynamically generate or adapt the feature matrix based on the data could reduce this limitation and allow the method to discover previously unknown or unexpected dynamics.
\end{enumerate}

Addressing these gaps will be crucial to extending the applicability and efficiency of the ARGOS method in more complex and real-world scenarios. The ARGOS method holds significant potential for various applications in power systems, particularly in the context of data-driven modeling and system identification:

\begin{enumerate}
    \item \textit{Grid Stability and Fault Detection}: ARGOS can be utilized to model the dynamic behavior of power grids during disturbances, enabling real-time identification of governing equations that describe system stability. By applying ARGOS to data collected from sensors and phasor measurement units, it could assist in detecting faults, predicting instabilities, and improving grid resilience.
    
    \item \textit{Renewable Energy Integration}: As the penetration of renewable energy sources such as wind and solar increases, ARGOS can help model the nonlinear dynamics of grid-following and grid-forming inverters, which are crucial for maintaining grid stability. These models can aid in optimizing the integration of renewables and ensuring their smooth operation within the existing power infrastructure.
    
    \item \textit{System Identification for Control Devices}: ARGOS can be applied to identify and model the dynamics of control devices used in modern power systems, such as inverters, converters, and storage systems. These models can improve the design and implementation of adaptive control strategies, leading to more efficient and responsive power system control.
    
    \item \textit{Dynamic Load Modeling}: Power system operators need accurate models of how loads respond to changes in grid conditions. ARGOS can be used to develop dynamic load models from real-time data, helping to improve the accuracy of load forecasts and contributing to better demand response management.
    
    \item \textit{Microgrid Optimization}: ARGOS can also be applied in the design and operation of microgrids, where accurate system identification is crucial for optimal energy management and control. By identifying governing equations that describe the behavior of various microgrid components, ARGOS can assist in improving the performance and reliability of these decentralized power systems.
\end{enumerate}

By leveraging its data-driven approach, ARGOS has the potential to revolutionize system identification in power systems, providing models that are accurate, concise, and interpretable, thus enabling more efficient and resilient grid operations.

\subsection{Deep SR}
\subsubsection{Strengths and Limitations}

The last method is deep SR, which involves utilizing deep learning techniques to uncover symbolic representations of data in the form of mathematical equations or formulas, boasts several advantages. Firstly, it offers interpretability, as symbolic representations are typically easier for humans to grasp compared to black-box deep learning models. This attribute is particularly valuable in fields where understanding the underlying mechanisms is crucial, such as scientific research and engineering. Secondly, deep SR models have the potential for strong generalization to unseen data, thanks to their focus on capturing fundamental mathematical relationships rather than merely memorizing specific data points. Moreover, deep SR automates the process of uncovering mathematical relationships within data, saving time and effort compared to manual exploration. Lastly, these models exhibit flexibility in capturing nonlinear relationships and intricate interactions within the data, rendering them applicable across various domains.

On the other hand, deep SR also presents several challenges and drawbacks. Firstly, training these models can be computationally demanding, especially when dealing with large datasets or complex symbolic representations. Secondly, there is a risk of overfitting, particularly when the model's complexity surpasses the amount of available training data. Additionally, deep SR models may struggle to capture certain types of relationships or patterns in the data, particularly those that are highly nonlinear or involve chaotic behavior. Furthermore, acquiring the substantial amount of labeled data required for training deep learning models may pose a challenge, especially in domains where data collection is resource-intensive. Another concern is the need for domain expertise to interpret the symbolic representations generated by these models, as the resulting equations or formulas may not be immediately intuitive to non-experts. Lastly, designing and implementing effective deep SR algorithms can be complex, necessitating expertise in both deep learning and symbolic computation. 

\subsubsection{Research Gaps and Potential Applications}
\label{sec:5C2}

Deep SR offers significant advantages such as interpretability and generalization capability across diverse datasets. These models excel in capturing nonlinear relationships and automating the discovery of fundamental mathematical structures from observational data, which is crucial in fields requiring transparent insights into underlying mechanisms. However, challenges include computational intensity, susceptibility to overfitting with limited data, and difficulties in handling highly complex or noisy datasets. Addressing these challenges through improved scalability, robustness to noise and outliers, and effective complexity control mechanisms are critical research areas. Enhancing these aspects would advance the practical applicability of deep SR in various domains, particularly where precise modeling and interpretative clarity are paramount. The specific research gaps that provide promising avenues for further exploration are discussed in details as follows:

\begin{enumerate}
 \item \textit{Scalability}:
 A significant research gap concerns scalability. Despite the effectiveness of deep learning techniques in learning complex representations from large datasets, they face substantial scalability challenges when confronted with intricate symbolic expressions or significant increases in dataset size. These challenges include issues such as computational efficiency, optimal memory usage, and the need for generalization across diverse datasets without excessive computational demands. Addressing these scalability concerns is crucial for advancing the practical applicability of deep SR, particularly in contexts requiring robust performance across extensive datasets.
 
 \item \textit{Robustness to Noise and Outliers}: 
 Another notable research gap pertains to robustness to noise and outliers. While deep learning methods excel in learning intricate patterns from data, they often struggle with maintaining performance in the presence of noisy or outlier-laden datasets. The challenge lies in ensuring that the learned symbolic representations are resilient to irrelevant or erroneous data points without compromising overall accuracy and generalization. Achieving robustness to noise and outliers is crucial for enhancing the reliability and practical utility of deep SR models, particularly in applications where data quality may vary significantly. Addressing this gap involves developing algorithms and techniques that can effectively filter out noise and mitigate the impact of outliers on the learned symbolic expressions.
 
 \item \textit{Complexity Control}: 
 A significant research gap concerns the issue of complexity control. While deep learning methodologies excel in capturing intricate data relationships, they often lack robust mechanisms for managing the complexity of generated symbolic expressions. This challenge is particularly pertinent in contexts where achieving a balance between model complexity and interpretability is paramount. Effective strategies for complexity control are crucial for tailoring symbolic representations to align with desired levels of complexity, thereby enhancing interpretative clarity, generalization capability, and computational efficiency. Addressing this gap necessitates the exploration and development of innovative techniques that can regulate and optimize model complexity while preserving predictive accuracy and interpretative fidelity.
 
\end{enumerate}

The identified research gaps in deep SR highlight critical areas for development that align with its promising applications in diverse domains. Improving scalability to manage complex symbolic expressions and large datasets is essential for optimizing control strategies and enhancing fault detection in power systems. Additionally, ensuring robustness to noise and outliers is crucial for accurate dynamic state estimation and parameter identification. Furthermore, effective complexity control mechanisms are necessary to balance model interpretability and computational efficiency in applications such as motor design and renewable energy forecasting. These applications leverage deep SR’s strengths in automation and flexibility to address operational challenges in highly complex systems. However, mitigating computational demands, overfitting risks, and enhancing interpretative clarity remains imperative through rigorous algorithmic advancements and empirical validation. These potential applications are discussed in details as follows:

\begin{enumerate} 

 \item \textit{Optimal Control}: Deep SR's ability to uncover interpretable symbolic representations can be leveraged to optimize control strategies for power electronics devices such as inverters, converters, and rectifiers. By discovering symbolic control laws, this approach can design more efficient and robust control algorithms that enhance the performance and stability of power systems, addressing its strength in capturing fundamental mathematical relationships.
 \item \textit{Fault Detection and Diagnosis}: The interpretability and scalability of deep SR make it suitable for analyzing operational data to identify patterns indicative of abnormalities or failures in power systems. This application benefits from this method's ability to handle large datasets and produce models that improve reliability and maintenance of power infrastructure, minimizing downtime and operational risks. 
 \item \textit{Dynamic State Estimation}: deep SR's flexibility in capturing nonlinear relationships can be utilized to estimate state variables of dynamic systems based on observed measurements. By analyzing historical data, deep SR models can infer current system states, such as voltage, current, power, and frequency in power systems. This application aligns with deep SR's strength in real-time monitoring, control, and optimization.
 \item \textit{Parameter Identification}: The interpretability and data-driven nature of deep SR can be employed to identify parameters of complex models representing power system components. By fitting symbolic equations to experimental or operational data, deep SR can estimate values like resistance, inductance, capacitance, and time constants, enabling accurate modeling and simulation, thereby leveraging its ability to automate mathematical relationship discovery.
 %\item \textit{Sensorless Control of Motors}: Applying deep SR to sensorless control of motors, including permanent magnet synchronous motors, switched reluctance motors, and induction motors, utilizes its strength in modeling and control through data-driven methods. This application highlights this method's ability to create models that do not rely on extensive sensor data, facilitating efficient control strategies.
 %\item \textit{Motor Design}: Optimizing motor design using deep SR involves leveraging its data-driven techniques for modeling and optimization. The method's scalability and flexibility help in discovering complex symbolic representations necessary for achieving desired performance metrics in motor design.
 \item \textit{Voltage and Frequency Regulation}: Deep SR techniques can develop models for voltage and frequency regulation in power systems, particularly in microgrids and distributed energy systems. By capturing dynamics of load variations, renewable energy generation, and energy storage devices, this method can facilitate the design of control strategies ensuring stable and reliable operation, utilizing its strength in handling intricate datasets.
 \item \textit{Renewable Energy Forecasting}: This method's potential for strong generalization to unseen data and its flexibility in capturing complex relationships can be applied to forecast renewable energy generation based on historical data and environmental variables. This application underscores deep SR's ability to handle large and intricate datasets to provide accurate predictions essential for grid integration and energy management.
 
\end{enumerate}

However, these applications also reflect deep SR's challenges. The computational demands and risk of overfitting, particularly in highly complex or noisy systems, necessitate careful model tuning and validation. Additionally, acquiring substantial labeled data and the need for domain expertise to interpret symbolic representations are critical considerations. Addressing these limitations through robust algorithm design and thorough experimentation is essential for these promising applications to be explored by researchers.

\subsection{Data Quality, Availability, and Volume}

SR methods rely heavily on the availability and quality of data. Power systems generate vast amounts of data from sensors, phasor measurement units (PMUs), and supervisory control and data acquisition (SCADA) systems. However, challenges such as missing data, noise, and inconsistent sampling rates can significantly impact model accuracy. Data sparsity in historical fault records and system disturbances further limits the ability to generalize SR models. Moreover, while high-dimensional data provide richer system dynamics, they pose computational and storage challenges that must be addressed through feature selection and dimensionality reduction techniques.

In addition to these challenges, power system data often come from heterogeneous sources, leading to variations in measurement accuracy and format. Ensuring data consistency across different measurement devices and operational conditions is crucial for maintaining model reliability. Data augmentation techniques, such as synthetic data generation and noise filtering, can be employed to enhance model performance in cases of insufficient or unreliable data. Furthermore, improving data-sharing protocols among utilities, research institutions, and industry stakeholders is essential for expanding access to high-quality datasets necessary for advancing SR methodologies.

\subsection{Real-World Validation of SR Models}

Despite the promising capabilities of SR in power systems, real-world validation remains a critical challenge. Many SR approaches are developed and validated on simulated datasets, which may not fully capture the complexities and uncertainties of actual grid operations. The deployment of SR-based models in real-world grids requires rigorous validation against field measurements to ensure robustness and reliability. Additionally, benchmarking against industry-standard methods, such as state estimation and physics-based dynamic models, is necessary to establish confidence in SR-derived equations. Uncertainties in system parameters, operational variability, and unseen disturbances must also be considered when assessing the generalization capability of SR models.

A significant challenge in real-world validation is the discrepancy between model assumptions and actual grid behavior. Power grids operate under dynamic conditions with evolving load profiles, renewable energy integration, and unforeseen disturbances. To ensure practical applicability, SR models must be tested under various operating scenarios, including contingency events and extreme weather conditions. Furthermore, collaboration with system operators and utility companies can facilitate access to real-world datasets, allowing for comprehensive testing and iterative refinement of SR models before deployment in critical decision-making processes.

\subsection{Deploying SR in Large-scale Power Grids}

The application of SR to large-scale power systems presents computational and practical challenges, particularly for real-time control and decision-making. The scalability of SR methods is hindered by the exponential increase in computational complexity as the number of state variables grows. Traditional SR approaches often struggle with real-time implementation due to the need for iterative optimization and model selection. Furthermore, integrating SR with control and optimization frameworks for power grid stability requires efficient and interpretable models that balance accuracy and computational efficiency.

One key challenge is the trade-off between model interpretability and complexity. While simpler models are easier to interpret and deploy, they may not capture the full nonlinear dynamics of large-scale grids. On the other hand, highly complex models can be computationally demanding, limiting their applicability in time-sensitive operations such as real-time monitoring and emergency response. To address these limitations, hybrid approaches that combine SR with machine learning, surrogate modeling, and high-performance computing techniques must be explored. Additionally, leveraging domain knowledge to constrain the search space can significantly improve computational feasibility for real-time applications. Developing hardware-accelerated solutions and parallel computing architectures can also enhance the scalability of SR-based models for large-scale power system applications.

\section{Conclusion}
\label{sec:6}
This study underscores the potential of SR methods, particularly SINDy, ARGOS, and deep SR, in advancing data-driven modeling in power systems. Each method demonstrates unique strengths: SINDy offers computational efficiency and interpretability, ARGOS excels in handling noisy data with confidence intervals, and deep SR integrates neural networks to capture complex nonlinearities. However, challenges persist in optimizing these models for high-dimensional, real-time applications. Future research should address these limitations by enhancing noise resilience, scalability, and model interpretability. Ultimately, SR methods provide powerful tools for identifying parsimonious models, contributing to improved control, optimization, and adaptability within modern power grids. This framework promises transformative impacts on the design and operational stability of dynamic power infrastructures, supporting the sustainable growth of energy systems worldwide.

\bibliographystyle{IEEEtran}
\bibliography{lib}

\end{document}